\documentclass[article]{IEEEtran}

\usepackage{color}
\usepackage{threeparttable}
\usepackage{srcltx}

\usepackage{graphicx,amsmath,psfrag,bm}
\usepackage[usenames,dvipsnames]{xcolor}
\usepackage{subfig}
\usepackage{cancel}
\usepackage{epstopdf}
\usepackage{pstool,cite}

\newcommand{\ve}[1]{\boldsymbol{#1}}
\newcommand{\te}[1]{\overline{\overline{#1}}}

\newcounter{tempEquationCounter}
\newcounter{thisEquationNumber}
\newenvironment{floatEq}
{\setcounter{thisEquationNumber}{\value{equation}}\addtocounter{equation}{1}
\begin{figure*}[!t]
\normalsize\setcounter{tempEquationCounter}{\value{equation}}
\setcounter{equation}{\value{thisEquationNumber}}
}
{\setcounter{equation}{\value{tempEquationCounter}}
\hrulefill\vspace*{4pt}
\end{figure*}
}

\begin{document}

\title{Design, Concepts and Applications of Electromagnetic Metasurfaces}
\author{Karim~Achouri and~Christophe~Caloz
\thanks{K. Achouri and C. Caloz are with the Department of Electrical Engineering, Polytechnique Montreal, Montreal, Quebec, Canada email: karim.achouri@polymtl.ca, christophe.caloz@polymtl.ca}%
}

\maketitle

\begin{abstract}
{The paper overviews our recent work on the synthesis of metasurfaces and related concepts and applications. The synthesis is based on generalized sheet transition conditions (GSTCs) with a bianisotropic surface susceptibility tensor model of the metasurface structure. We first place metasurfaces in a proper historical context and describe the GSTC technique with some fundamental susceptibility tensor considerations. Upon this basis, we next provide an in-depth development of our susceptibility-GSTC synthesis technique. Finally, we present five recent metasurface concepts and applications, which cover the topics of birefringent transformations, bianisotropic refraction, light emission enhancement, remote spatial processing and nonlinear second-harmonic generation.}
\end{abstract}

\section{Introduction}
\label{sec:intro}

Metamaterials reached a peak of interest in the first decade of the 21st century. Then, due to their fabrication complexity, bulkiness and weight, and their limitations in terms of losses, frequency range and scalability, they became less attractive, and were progressively superseded by their  two-dimensional counterparts, the metasurfaces~\cite{Holloway2009,holloway2012overview,Minovich2015,Glybovski20161}.

\begin{figure}[h!]
\begin{center}
\subfloat[]{\label{fig:FZPR}
\includegraphics[width=0.5\columnwidth]{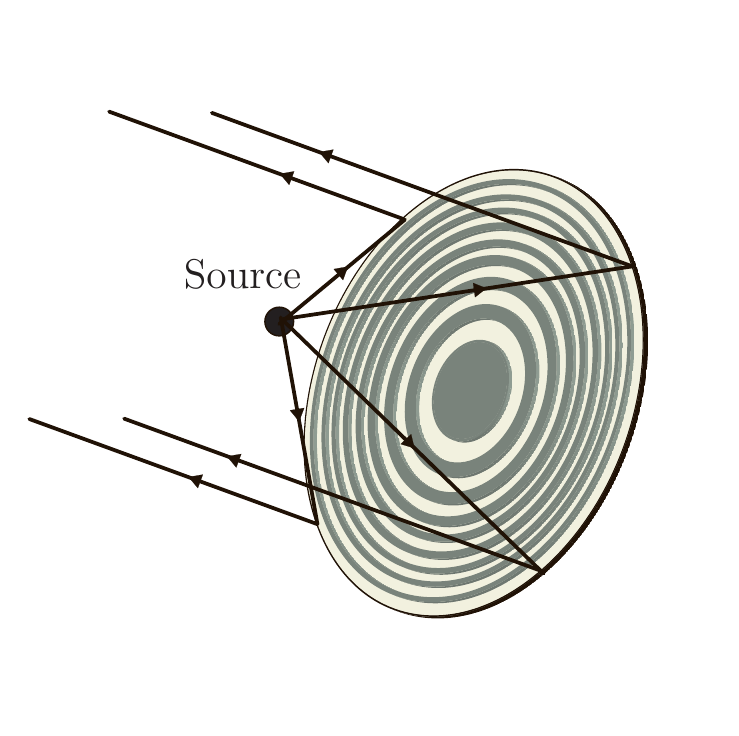}
}
\subfloat[]{\label{fig:FSS}
\includegraphics[width=0.5\columnwidth]{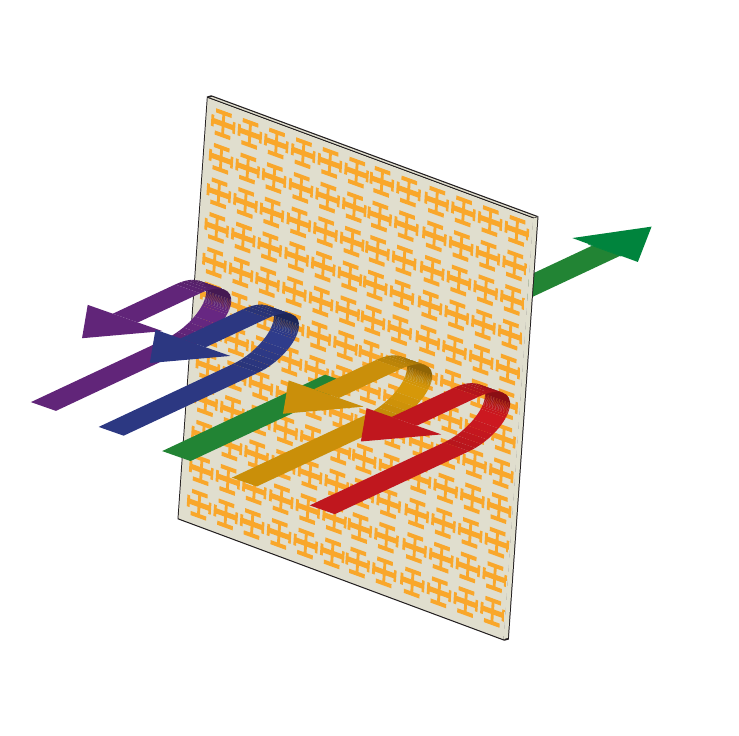}
}\\
\subfloat[]{\label{fig:Reflarray}
\includegraphics[width=0.5\columnwidth]{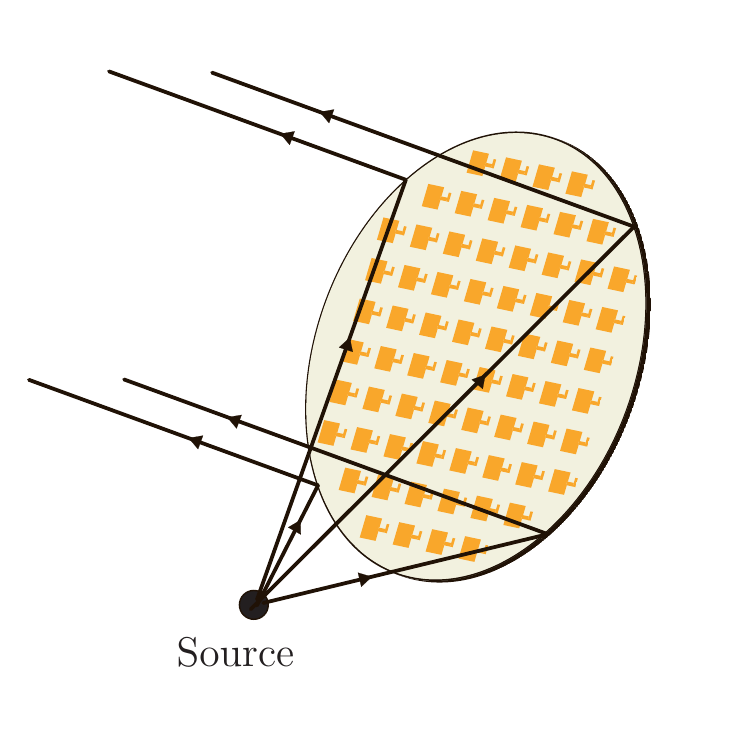}
}
\subfloat[]{\label{fig:Arraylens}
\includegraphics[width=0.5\columnwidth]{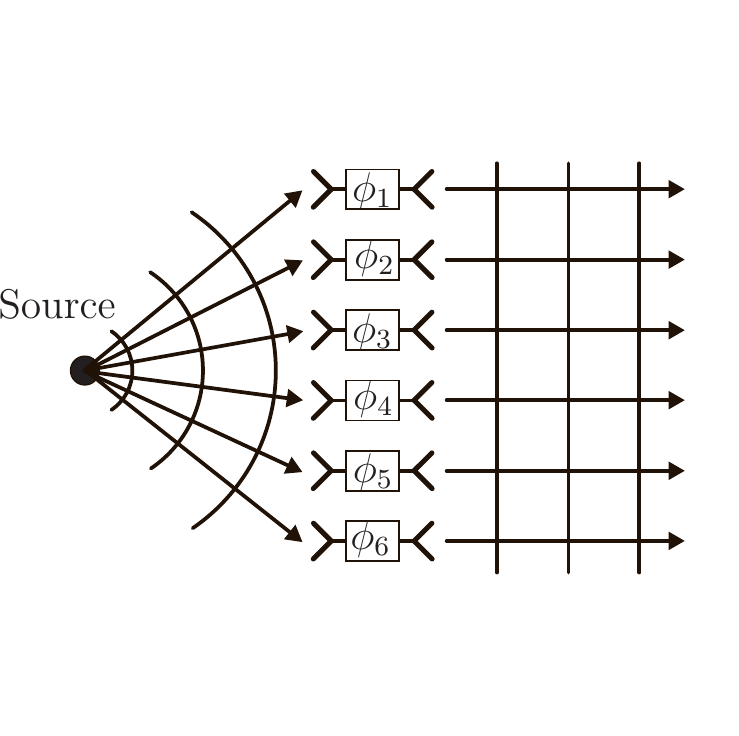}
}
\caption{Examples of two-dimensional wave manipulating structures: (a)~Fresnel zone plate reflector, (b)~reflectarray, (c)~interconnected array lens and (d)~frequency-selective surface.}
\label{fig:twoD}
\end{center}
\end{figure}

The idea of controlling electromagnetic waves with electromagnetically thin structures is clearly not a new concept. The first example is probably that of Lamb, who studied the reflection and transmission from an array of metallic strips, already back in 1897~\cite{lamb1897reflection}. Later, in the 1910s, Marconi used arrays of straight wires to realize polarization reflectors~\cite{marconi1919reflector}. These first two-dimensional electromagnetic structures were later followed by a great diversity of systems that emerged mainly with the developments of the radar technology during World War II. Many of these systems date back to the 1960s. The Fresnel zone plate reflectors, illustrated in Fig.~\ref{fig:FZPR}, were based on the concept of the Fresnel lens demonstrated almost 150 years earlier and used in radio transmitters~\cite{1144988}. The frequency selective surfaces (FSS), illustrated in Fig.~\ref{fig:FSS}, were developed as spatial filters~\cite{MunkFSS,6867682}. The reflectarray antennas~\cite{huang2007reflectarray}, were developed as the flat counterparts of parabolic reflectors, and were initially formed by short-ended waveguides~\cite{1138112}.
They were later progressively improved and the short-ended waveguides were replaced with mircrostrip printable scattering elements in the late 1970s~\cite{malagisi1978microstrip,montgomery1978microstrip}, as shown in Fig.~\ref{fig:Reflarray}. The transmissive counterparts of the reflectarrays are the transmitarrays, which were used as array lens systems and date back to the 1960s~\cite{1150235,1448703,1143726}. They were first implemented in the form of two interconnected planar arrays of dipole antennas, one for receiving and one for transmitting, where each antenna on the receiver side was connected via a delay line to an antenna on the transmit side, as depicted in Fig.~\ref{fig:Arraylens}. Through the 1990ies, the transmitarrays evolved from interconnected antenna arrays to layered metallic structures that were essentially the functional extensions of FSS~\cite{543779,659416,959839} with efficiency limited by the difficulty to control the transmission phase over a $2\pi$ range while maintaining a high enough amplitude. Finally, compact quasi-transparent transmitarrays or phase-shifting surfaces, able to cover a $2\pi$-phase range, were demonstrated in 2010~\cite{5395659}.

The aforementioned Fresnel lenses, FSS, reflectarrays and transmitarrays are the precursors of today's \emph{metasurfaces}\footnote{So far, and throughout this paper, we essentially consider metasurfaces illuminated by waves incident on the them under a non-zero angle with respect to their plane, i.e. space waves, which represent the metasurfaces leading to the main applications. However, metasurface may also be excited within their plane, i.e. by surface waves or leaky-waves, as in~\cite{Imani2010104,Grbic25042008,6576211,5498959,6127895}.}.

From a general perspective, metasurfaces can be used to manipulate the polarization, the phase and the amplitude of electromagnetic fields. A rich diversity of metasurface applications have been reported in the literature to date and many more are expected to emerge. These applications are too numerous to be exhaustively cited. Some of the most significant ones are reported in~\cite{1.4820810,6704736,6692975,2367668,PhysRevX.4.021026,Huang2013345,MOP:MOP29003,4869127, Kenanakis:12}  (polarization tranformations), \cite{li2014ultra,4832785,PhysRevB.91.115305,Wen:14,dincer2014polarization,6748871,6553200} (absorption) and~\cite{capasso1,yu2014flat,6748002,Shi:14,Yang:14,PhysRevLett.114.095503,Lipworth:13,Hunt18012013,6933919,4821357,20060309,6805160,6705631,6894567}  (wavefront manipulations). More sophisticated metasurfaces, transforming both phase and polarization, have been recently realized. This includes metasurfaces producing beams possessing angular orbital momentum~\cite{nye1974dislocations} or vortex waves~\cite{karimi2014generating,PhysRevApplied.2.044012,Yi:14,Veysi:15,nl500658n,doi:10.1021/nl402039y}, holograms~\cite{Hunt:14,zheng2015metasurface} and stable beam traction~\cite{PhysRevB.91.115408}. Additionally, nonreciprocal transformations~\cite{PhysRevB.88.205405,PhysRevE.89.053203,PhysRevB.89.075109,3615688,6280630}, nonlinear interactions~\cite{1.4914343,valev2014nonlinear,lee2014giant}, analog computing~\cite{nl5047297,Silva10012014} and spatial filtering~\cite{ortiz2013spatial,PhysRevB.90.125422,Shen28032014} have also been reported.

To deploy their full potential, metasurfaces must be designed efficiently. This requires a solid \emph{model}, that both simplifies the actual problem and provides insight into its physics. Metasurfaces are best modelled, according to Huygens' principle, as \emph{surface polarization current sheets} via continuous (locally homogeneous) bianisotropic surface susceptibility tensorial functions. Inserting the corresponding surface polarization densities into Maxwell equations results in electromagnetic sheet transition conditions, which consist in the key equations to solve in the design of metasurfaces.

The objective of this paper is twofold. First, it will present a general framework for the synthesis of the aforementioned metasurface surface susceptibility functions for arbitrary (amplitude, phase, polarization, propagation direction and waveform) specified fields. From this point, the physical structure (material and geometry of the scattering particles, substrate parameters and layer configuration, thickness and size) is tediously but straightforwardly determined, after discretization of the susceptibility functions, using scattering parameter mapping. The synthesis of metasurfaces has been the objective of many researches in recent years~\cite{achouri2014general,salem2014metasurface,Salem2013c,nl5001746,Pfeiffer2013a,6648706,PhysRevApplied.2.044011,Wong2014360,6905746,6891256,6477089,452013}. Second, the paper will show how this synthesis framework provides a general perspective of the electromagnetic transformations achievable by metasurfaces, and present subsequent concepts and applications.

\section{Sheet Transition Conditions}
\label{sec:BC}

\begin{figure}[ht]
\centering
\includegraphics[width=0.8\linewidth]{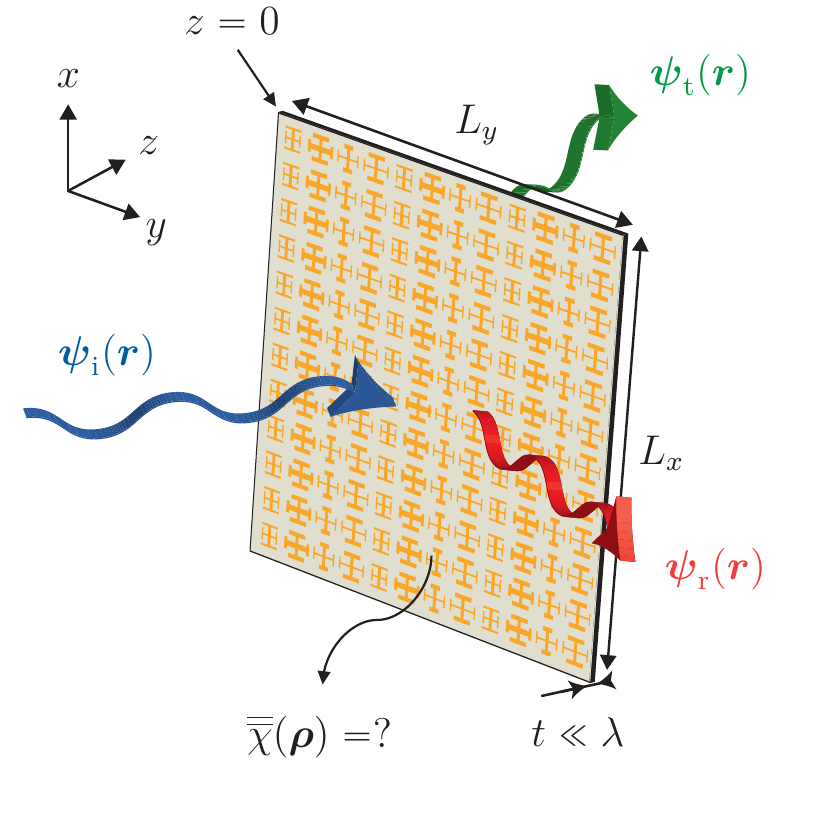}
\caption{Metasurface synthesis problem. The metasurface to be synthesized lies in the $xy$-plane at $z=0$. The synthesis procedure consists in finding the susceptibility tensors characterizing the metasurface, $\te{\chi}(\ve{\rho})$, in terms of specified arbitrary incident, $\ve{\psi}_\text{i}(\ve{r})$, reflected, $\ve{\psi}_\text{r}(\ve{r})$, and transmitted, $\ve{\psi}_\text{t}(\ve{r})$, waves.}
\label{fig:MS}
\end{figure}

The general synthesis problem of a metasurface is represented in Fig.~\ref{fig:MS}. As mentioned in Sec.~\ref{sec:intro}, the metasurface is modeled as an electromagnetic sheet (zero-thickness film)\footnote{This approximation is justified by the fact that a physical metasurface is electromagnetically very thin, so that it cannot support significant phase shifts and related effects, such as Fabry-Perot resonances.}. In the most general case, a metasurface is made of an array of polarizable scattering particles that induce both electric and magnetic field discontinuities. It is therefore necessary to express the discontinuities of these fields as functions of the electric and magnetic surface polarization densities ($\ve{P}$ and $\ve{M}$). The rigorous boundary conditions that apply to such an interface have been originally derived by Idemen~\cite{Idemen1973}.

For a metasurface lying in the $xy$-plane at $z=0$, these transition conditions follow from the idea that all the quantities in Maxwell equations can be expressed in the following form
\begin{equation}
\label{eq:General_F}
f(z)=\left \{ f(z) \right\}+\sum_{k=0}^{N}f_{k}\delta^{(k)}(z),
\end{equation}
where the function $f(z)$ is discontinuous at $z=0$. The first term of the right-hand side of~\eqref{eq:General_F} is the \emph{regular part} of $f$, which corresponds to the value of the function everywhere except at $z=0$, while the second term is the \emph{singular part} of $f$, which is an expansion over the $k$-th derivatives of the Dirac delta distribution (corresponding to the discontinuity of $f$ and the $k$-th derivatives of $f$).

Most often, the series in~\eqref{eq:General_F} may be truncated at $N=0$, so that only the discontinuities of the fields are taken into account while the discontinuities of the derivatives of the fields are neglected. With this truncation, the metasurface transition conditions, known as the generalized sheet transition conditions (GSTCs), are found as\footnote{Note that these relations can also be obtained following the more traditional technique of box integration, as demonstrated in~\cite{albooyeh2016electromagnetic}.}

\begin{subequations}
\label{eq:BC}
\begin{align}
\ve{\hat{z}}\times\Delta\ve{H}
&=j\omega\ve{P}_{\parallel}-\ve{\hat{z}}\times\nabla_{\parallel}M_{z},\label{eq:CurlH}\\
\Delta\ve{E}\times\ve{\hat{z}}
&=j\omega\mu_0 \ve{M}_{\parallel}-\nabla_{\parallel}\bigg(\frac{P_{z}}{\epsilon_0 }\bigg)\times\ve{\hat{z}},\label{eq:CurlE}\\
\ve{\hat{z}}\cdot\Delta\ve{D}
&=-\nabla\cdot\ve{P}_{\parallel},\label{eq:divD}\\
\ve{\hat{z}}\cdot\Delta\ve{B}
&=-\mu_0 \nabla\cdot\ve{M}_{\parallel},\label{eq:divB}
\end{align}
\end{subequations}
where the terms on the left-hand sides of the equations correspond to the differences of the fields on both sides of the metasurface, which may be expressed as
\begin{equation}
\label{eq:field_diff}
\Delta \Psi_{u}
=\ve{\hat{u}}\cdot\Delta\ve{\Psi}\Bigr|_{z=0^{-}}^{0^{+}}
=\Psi_{u,\text{t}}-(\Psi_{u,\text{i}}+\Psi_{u,\text{r}}),\; u=x,y,z,
\end{equation}
where $\ve{\Psi}$ represents any of the fields $\ve{E}$, $\ve{H}$, $\ve{D}$ or $\ve{B}$, and where the subscripts i, r, and t denote the incident, reflected and transmitted fields, and $\ve{P}$ and $\ve{M}$ are the electric and magnetic surface polarization densities, respectively.

In the general case of a linear bianisotropic metasurface, these polarization densities are related to the acting (or local) fields, $\ve{E}_\text{act}$ and $\ve{H}_\text{act}$, by~\cite{kong1986electromagnetic,lindell1994electromagnetic}
\begin{subequations}
\label{eq:pola_dens}
\begin{align}
\ve{P}&=\epsilon_0 N\te{\alpha}_{\text{ee}}\cdot\ve{E}_\text{act}+\frac{1}{c_0} N\te{\alpha}_{\text{em}}\cdot\ve{H}_\text{act},\\
\ve{M}&=N\te{\alpha}_{\text{mm}}\cdot\ve{H}_\text{act}+\frac{1}{\eta_0}N\te{\alpha}_{\text{mm}}\cdot\ve{E}_\text{act},
\end{align}
\end{subequations}
\noindent where the $\te{\alpha}_{\text{ab}}$ tensors represent the polarizabilities of a given scatterer, $N$ is the number of scatterers per unit area, $c_0$ is the speed of light in vacuum and $\eta_0$ is the vacuum impedance\footnote{Despite being indeed quite general, these relations are still restricted to \emph{linear} and \emph{time-invariant} metasurfaces. The synthesis of nonlinear metasurfaces has been approached using extended GSTCs in~\cite{achouri2017mathematical}.}. This is a \emph{microscopic} description of the metasurface response which requires an appropriate definition of the coupling between adjacent scattering particles. In this work, we use the concept of susceptibilities rather than the polarizabilities to provide a \emph{macroscopic} description of the metasurface, which allows a direct connection with material parameters such as $\te{\epsilon}_r$ and $\te{\mu}_r$. To bring about the susceptibilities, relations~\eqref{eq:pola_dens} can be transformed by noting that the acting fields, at the position of a scattering particle, can be defined as the average total fields minus the field scattered by the considered particle~\cite{kuester2003av}, i.e. $\ve{E}_\text{act}=\ve{E}_\text{av} - \ve{E}_\text{scat}^\text{part}$. The contributions of the particle may be expressed by considering the particle as a combination of electric and magnetic dipoles contained within a small disk, and the field scattered from this disk can be related to $\ve{P}$ and $\ve{M}$ by taking into account the coupling with adjacent scattering particles. Therefore, the acting fields are functions of the average fields and the polarization densities. Upon substitution of this definition of the acting fields in~\eqref{eq:pola_dens}, the expressions of the polarization densities become
\begin{subequations}
\label{eq:pola_dens_2}
\begin{align}
 \ve{P}&=\epsilon_0 \te{\chi}_\text{ee}\cdot\ve{E}_\text{av}+\frac{1}{c_0}\te{\chi}_\text{em}\cdot\ve{H}_\text{av},\\
 \ve{M}&=\te{\chi}_\text{mm}\cdot\ve{H}_\text{av}+\frac{1}{\eta_0}\te{\chi}_\text{me}\cdot\ve{E}_\text{av},
\end{align}
\end{subequations}
where the average fields are defined as
\begin{equation}
\label{eq:field_av}
\Psi_{u,\text{av}}
=\ve{\hat{u}}\cdot\ve{\Psi}_\text{av}
=\frac{\Psi_{u,\text{t}}+(\Psi_{u,\text{i}}+\Psi_{u,\text{r}})}{2},
\; u=x,y,z,
\end{equation}
where $\ve{\Psi}$ corresponds to $\ve{E}$ or $\ve{H}$.

\section{Susceptibility Tensor Considerations}

Before delving into the metasurface synthesis, it is important to examine the susceptibility tensors in~\eqref{eq:pola_dens_2} in the light of fundamental electromagnetic considerations pertaining to reciprocity, passivity and loss.

The \emph{reciprocity} conditions for a bianisotropic metasurface, resulting from the Lorentz theorem~\cite{kong1986electromagnetic}, read
\begin{equation}
\label{eq:reciprocity}
\te{\chi}_{\text{ee}}^{\text{T}} = \te{\chi}_{\text{ee}}, \quad \te{\chi}_{\text{mm}}^{\text{T}}=\te{\chi}_{\text{mm}}, \quad \te{\chi}_{\text{me}}^{\text{T}}=-\te{\chi}_{\text{em}},
\end{equation}
where the superscripts $T$ denotes the matrix transpose operation\footnote{These conditions are identical to those for a bianisotropic medium~\cite{kong1986electromagnetic,lindell1994electromagnetic}, except that the metasurfaces in~\eqref{eq:reciprocity} are surface instead of volume susceptibilities}.

Adding the property of losslessness, resulting from the bianisotropic Poynting theorem~\cite{kong1986electromagnetic}, restricts~\eqref{eq:reciprocity} to
\begin{equation}
\label{eq:lossless}
\te{\chi}_{\text{ee}}^{\text{T}} = \te{\chi}_{\text{ee}}^\ast, \quad \te{\chi}_{\text{mm}}^{\text{T}}=\te{\chi}_{\text{mm}}^*, \quad \te{\chi}_{\text{me}}^{\text{T}}=\te{\chi}_{\text{em}}^*,
\end{equation}
which characterize a simultaneously passive, lossless and reciprocal metasurface.

The conditions~\eqref{eq:reciprocity} and~\eqref{eq:lossless} establish relations between different susceptibility components of the constitutive tensors. Therefore, requiring the metasurface to be reciprocal or reciprocal and lossless/gainless, as often practically desirable, reduces the number of independent susceptibility components~\cite{achouri2014general,Achouri2015c,Achouri2015b}, and hence reduces the diversity of achievable field transformations, as will be shown next.

\section{Metasurface Synthesis}
\label{sec:Syn}

\subsection{General Concepts}
\label{sec:GenConc}

We follow here the metasurface synthesis procedure\footnote{The \emph{synthesis} procedure consists in determining the physical metasurface structure for specified fields. The inverse procedure is the \emph{analysis}, which consists in determining the fields scattered by a given physical metasurface structures for a given incident field, and is generally coupled (typically iteratively) with the synthesis for the efficient design of a metasurface~\cite{Vahab2017metasurface}. The overall design procedure thus consists of the combination of the synthesis and analysis operations. This paper focuses on the direct synthesis of the susceptibility functions, as this is the most important aspect for the understanding of the physical properties of metasurfaces, the elaboration of related concepts, and the development of resulting applications.} introduced in~\cite{achouri2014general}, which seems the most general approach reported to date. This procedure consists in solving the GSTCs equations~\eqref{eq:BC} to determine the unknown susceptibilities in~\eqref{eq:pola_dens_2} required for the metasurface to perform the electromagnetic transformation specified in terms of the incident, reflected and transmitted fields. Note that Eqs.~\eqref{eq:divD} and~\eqref{eq:divB} are redundant in the system~\eqref{eq:BC}, due to the absence of impressed sources, so that Eqs.~\eqref{eq:CurlH} and~\eqref{eq:CurlE} are sufficient to fully describe the metasurface and synthesize it. Consequently, only the transverse (tangential to the metasurface) components of the specified fields, explicitly apparent in~\eqref{eq:field_diff} and in~\eqref{eq:field_av}, are involved in the the synthesis, even though these fields may generally include longitudinal (normal to the metasurface) components as well. According to the uniqueness theorem, the longitudinal components of the fields are automatically determined from the transverse fields.

The GSTC equations~\eqref{eq:CurlH} and~\eqref{eq:CurlE} form a set of (inhomogeneous) coupled  partial differential equations, due to the partial derivatives of the normal components of the polarization densities, $P_{z}$ and $M_{z}$. The resolution of the corresponding inverse problem is nontrivial and requires involved numerical processing. In contrast, if $P_{z}=M_{z}=0$, the differential system reduces to a simple algebraic system of equations, most conveniently admitting closed-form solutions for the synthesized susceptibilities. For this reason, we will focus on this case in this section, while a transformation example with nonzero normal susceptibilities will be discussed in Sec.~\ref{sec:PzMz}.

Enforcing that $P_{z}=M_{z}=0$ may a priori seem to represent an important restriction, particularly, as we shall see, in the sense that it reduces the number of degrees of freedom of the metasurface. However, this is not a major restriction since a metasurface with normal polarization currents can generally be reduced to an equivalent metasurface with purely tangential polarization currents, according to Huygens' theorem. This restriction mostly affects the realization of the scattering particles, that are then forbidden to exhibit normal polarizations, which ultimately limits their practical implementation\footnote{Moreover, in the particular case where all the specified waves are normal to the metasurface, the excitation of normal polarization densities do not induce any discontinuity in the fields. This is because the corresponding fields, and hence the related susceptibilities, are not functions of the $x$ and $y$ coordinates, so that the spatial derivatives of $P_z$ and $M_z$ in Eqs.~\eqref{eq:CurlH} and~\eqref{eq:CurlE} are zero, i.e. do not induce any discontinuity in the fields across the metasurface. Thus, susceptibilities producing normal polarizations can be ignored, and only tangential susceptibility components must be considered, when the metasurface is synthesized for normal waves.}.

Substituting the constitutive relations~\eqref{eq:pola_dens_2} into the GSTCs~\eqref{eq:CurlH} and \eqref{eq:CurlE} with $M_z=P_z=0$ leads to
\begin{subequations}
\label{eq:InvProb}
\begin{align}
\ve{\hat{z}}\times\Delta\ve{H}
&=j\omega\epsilon_0\te{\chi}_\text{ee}\cdot\ve{E}_\text{av}+jk_0\te{\chi}_\text{em}\cdot\ve{H}_\text{av},\label{eq:diffH}\\
\Delta\ve{E}\times\ve{\hat{z}}
&=j\omega\mu_0 \te{\chi}_\text{mm}\cdot\ve{H}_\text{av}+jk_0\te{\chi}_\text{me}\cdot\ve{E}_\text{av},\label{eq:diffE}
\end{align}
\end{subequations}
where $k_0=\omega/c_0$ is the free-space wavenumber and where the susceptibility tensors only contain the tangential susceptibility components. This system can also be written in matrix form
\begin{equation}
\label{eq:InvProbMatrix}
\begin{pmatrix}
\Delta H_y\\
\Delta H_x\\
\Delta E_y\\
\Delta E_x
\end{pmatrix}=
\begin{pmatrix}
\widetilde{\chi}_\text{ee}^{xx} & \widetilde{\chi}_\text{ee}^{xy} & \widetilde{\chi}_\text{em}^{xx} & \widetilde{\chi}_\text{em}^{xy}\\
\widetilde{\chi}_\text{ee}^{yx} & \widetilde{\chi}_\text{ee}^{yy} & \widetilde{\chi}_\text{em}^{yx} & \widetilde{\chi}_\text{em}^{yy}\\
\widetilde{\chi}_\text{me}^{xx} & \widetilde{\chi}_\text{me}^{xy} & \widetilde{\chi}_\text{mm}^{xx} & \widetilde{\chi}_\text{mm}^{xy}\\
\widetilde{\chi}_\text{me}^{yx} & \widetilde{\chi}_\text{me}^{yy} & \widetilde{\chi}_\text{mm}^{yx} & \widetilde{\chi}_\text{mm}^{yy}
\end{pmatrix}
\cdot
\begin{pmatrix}
E_{x,\text{av}}\\
E_{y,\text{av}}\\
H_{x,\text{av}}\\
H_{y,\text{av}}
\end{pmatrix},
\end{equation}
where the tilde symbol indicates normalized susceptibilities, related to the non-normalized susceptibilities in~\eqref{eq:InvProb} by
\begin{equation}
\label{eq:conv}
\begin{split}
&
\begin{pmatrix}
\chi_\text{ee}^{xx} & \chi_\text{ee}^{xy} & \chi_\text{em}^{xx} & \chi_\text{em}^{xy}\\
\chi_\text{ee}^{yx} & \chi_\text{ee}^{yy} & \chi_\text{em}^{yx} & \chi_\text{em}^{yy}\\
\chi_\text{me}^{xx} & \chi_\text{me}^{xy} & \chi_\text{mm}^{xx} & \chi_\text{mm}^{xy}\\
\chi_\text{me}^{yx} & \chi_\text{me}^{yy} & \chi_\text{mm}^{yx} & \chi_\text{mm}^{yy}
\end{pmatrix}=\\
&\qquad
=\begin{pmatrix}
\frac{j}{\omega\epsilon_0}\widetilde{\chi}_\text{ee}^{xx} & \frac{j}{\omega\epsilon_0}\widetilde{\chi}_\text{ee}^{xy} & \frac{j}{k_0}\widetilde{\chi}_\text{em}^{xx} & \frac{j}{k_0}\widetilde{\chi}_\text{em}^{xy}\\
-\frac{j}{\omega\epsilon_0}\widetilde{\chi}_\text{ee}^{yx} & -\frac{j}{\omega\epsilon_0}\widetilde{\chi}_\text{ee}^{yy} & -\frac{j}{k_0}\widetilde{\chi}_\text{em}^{yx} & -\frac{j}{k_0}\widetilde{\chi}_\text{em}^{yy}\\
-\frac{j}{k_0}\widetilde{\chi}_\text{me}^{xx} & -\frac{j}{k_0}\widetilde{\chi}_\text{me}^{xy} & -\frac{j}{\omega\mu_0}\widetilde{\chi}_\text{mm}^{xx} & -\frac{j}{\omega\mu_0}\widetilde{\chi}_\text{mm}^{xy}\\
\frac{j}{k_0}\widetilde{\chi}_\text{me}^{yx} & \frac{j}{k_0}\widetilde{\chi}_\text{me}^{yy} & \frac{j}{\omega\mu_0}\widetilde{\chi}_\text{mm}^{yx} & \frac{j}{\omega\mu_0}\widetilde{\chi}_\text{mm}^{yy}
\end{pmatrix}.
\end{split}
\end{equation}
The system~\eqref{eq:InvProbMatrix} contains 4 equations for 16 unknown susceptibilities. It is therefore heavily under-determined and cannot be solved directly\footnote{Even if it would be solved, this would probably result in an inefficient metasurface, as it would use more susceptibility terms than required to accomplish the specified task.}. This leaves us with two distinct resolution possibilities.

The first possibility would be to reduce the number of susceptibilities from 16 to 4 in order to obtain a fully determined (full-rank) system. Since there exists many combinations of susceptibility quadruplets\footnote{Mathematically, the number of combinations would be $16!/[(16-4)!4!]=1,820$, but only a subset of these combinations represent physically meaningful combinations.}, different sets can be chosen, each of them naturally corresponding to different field transformations. This approach thus requires an educated selection of the susceptibility quadruplet that is the most likely to enable the specified operation, within existing constraints\footnote{For instance, the specification of a reciprocal transformation, corresponding to the metasurface properties in Eq.~\eqref{eq:reciprocity}, would automatically preclude the selection of off-diagonal pairs for $\te{\chi}_{\text{ee,mm}}$.}.

These considerations immediately suggest that a second possibility would be to augment the number of field transformation specifications, i.e. allow the metasurface to perform more independent transformations, which may be of great practical interest in some applications. We would have thus ultimately three possibilities to resolve~\eqref{eq:InvProbMatrix}: a)~reducing the number of independent unknowns, b)~increasing the number of transformations and c)~a combination of a) and b).

As we shall see in the forthcoming sections, the number ${\cal N}$ of physically or practically achievable transformations for a metasurface with $P$ susceptibility parameters, ${\cal N}(P)$, is not trivial; specifically, ${\cal N}(P)=P/4$, that may be expected from a purely mathematical viewpoint, is not always true!

\subsection{Four-Parameter Transformation}
\label{sec:singletrans}

We now provide an example for the approach where the number of susceptibility parameters has been reduced to 4, or $P=4$, so that the system~\eqref{eq:InvProbMatrix} is of full-rank nature. We thus have to select 4 susceptibility parameters and set all the others to zero in~\eqref{eq:conv}. We decide to consider the simplest case of a monoanisotropic (8 parameters $\widetilde{\chi}_\text{em,me}^{uv}=0$, $u,v=x,y$) axial (4 parameters $\widetilde{\chi}_\text{ee,mm}^{uv}=0$ for $u\neq v$, $u,v=x,y$) metasurface, which is thus characterized by the four parameters $\widetilde{\chi}_\text{ee}^{xx}$, $\widetilde{\chi}_\text{ee}^{yy}$, $\widetilde{\chi}_\text{mm}^{xx}$ and $\widetilde{\chi}_\text{mm}^{yy}$, so that Eq.~\eqref{eq:InvProbMatrix} reduces to the diagonal system
\begin{equation}
\label{eq:birefsystem}
\begin{pmatrix}
\Delta H_y\\
\Delta H_x\\
\Delta E_y\\
\Delta E_x
\end{pmatrix}=
\begin{pmatrix}
\widetilde{\chi}_\text{ee}^{xx} & 0 & 0 & 0 \\
0 & \widetilde{\chi}_\text{ee}^{yy} & 0 & 0 \\
0 & 0 & \widetilde{\chi}_\text{mm}^{xx} & 0 \\
0 & 0 & 0 & \widetilde{\chi}_\text{mm}^{yy}
\end{pmatrix}
\cdot
\begin{pmatrix}
E_{x,\text{av}}\\
E_{y,\text{av}}\\
H_{x,\text{av}}\\
H_{y,\text{av}}
\end{pmatrix}.
\end{equation}
This metasurface is a \emph{biregringent} structure~\cite{saleh2007fundamentals}, with decoupled $x$-polarized and $y$-polarized susceptibility pairs
\begin{subequations}
\label{eq:chi_diag}
\begin{equation}
\chi_{\text{ee}}^{xx}=\frac{j\Delta H_{y}}{\omega\epsilon_0  E_{x,\text{av}}},
\quad\chi_{\text{mm}}^{yy}=\frac{j\Delta E_{x}}{\omega\mu_0  H_{y,\text{av}}}\label{eq:chi_diag_Exx_Myy}
\end{equation}
and
\begin{equation}
\chi_{\text{ee}}^{yy}=\frac{-j\Delta H_{x}}{\omega\epsilon_0  E_{y,\text{av}}},
\quad\chi_{\text{mm}}^{xx}=\frac{-j\Delta E_{y}}{\omega\mu_0  H_{x,\text{av}}},\label{eq:chi_diag_Eyy_Mxx}
\end{equation}
\end{subequations}
respectively\footnote{If the two electric and the two magnetic susceptibilities in~\eqref{eq:chi_diag} are equal to each other ($\chi_{\text{ee}}^{xx} = \chi_{\text{ee}}^{yy}$ and $\chi_{\text{mm}}^{xx} = \chi_{\text{mm}}^{yy}$), the monoanisotropic metasurface in~\eqref{eq:birefsystem} reduces to the simplest possible case of a monoisotropic metasurface, and hence performs the same operation for $x$- and $y$-polarized waves.}. In these relations, according to~\eqref{eq:field_diff} and~\eqref{eq:field_av}, $\Delta H_y=H_{y,\text{t}}-(H_{y,\text{i}}+H_{y,\text{r}})$, $E_{x,\text{av}}=(E_{x,\text{t}}+E_{x,\text{i}}+E_{x,\text{r}})/2$, and so on. By synthesis, the metasurface with the susceptibilities~\eqref{eq:chi_diag} will exactly transform the specified incident field into the specified reflected and transmitted fields, in an arbitrary fashion, except for the constraint of reciprocity since the susceptibility tensor in~\eqref{eq:birefsystem} inherently satisfies~\eqref{eq:reciprocity}.

It should be noted that the example of~\eqref{eq:birefsystem}, with 4 distinct susceptibility parameters, is a very particular case of a four-parameter transformation since the components in~\eqref{eq:chi_diag_Exx_Myy} and~\eqref{eq:chi_diag_Eyy_Mxx} are decoupled from each other, which is the origin of birefringence. Now, birefringence may be considered as a \emph{pair} of distinct and independent transformations (one for $x$-polarization one for $y$-polarization), i.e. ${\cal N}(4)=2>4/4$. Thus, the specification of 4 susceptibility parameters may lead to more than 1 transformation, which, by extension, already suggests that $P$ susceptibilities may lead to more than $P/4$ transformations, as announced in Sec.~\ref{sec:GenConc} and will be further discussed in Sec.~\ref{sec:MultiTrans}.

\begin{figure}[ht]
\centering
\includegraphics[width=1\linewidth]{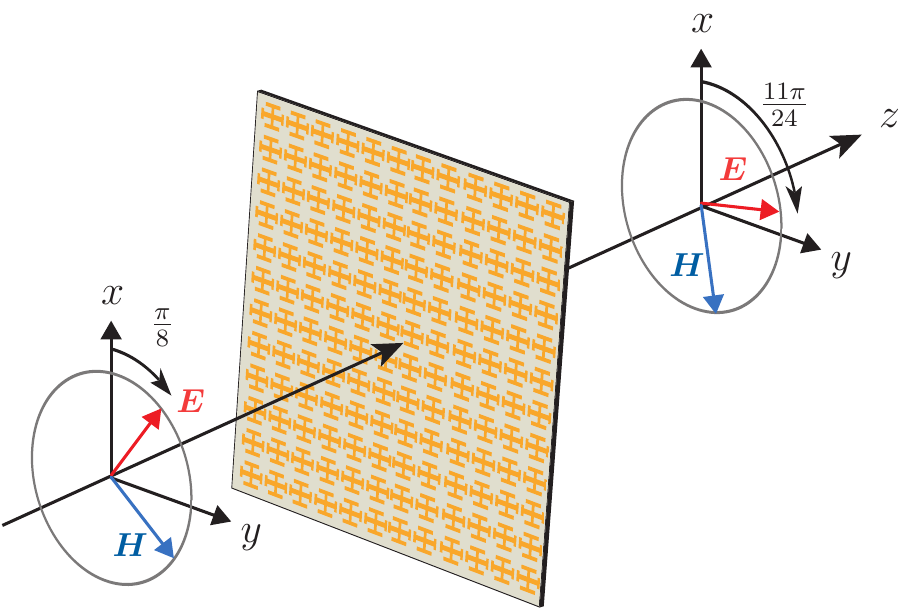}
\caption{Polarization reflectionless rotating metasurface. The metasurface rotates the polarization of a linearly polarized normally incident plane wave from the angle $\pi/8$ to the angle $11\pi/24$ with respect to the $x$-axis (rotation of $\pi/3$). The metasurface is surrounded on both sides by vacuum, i.e. $\eta_1 = \eta_2 = \eta_0$.}
\label{fig:Faraday_rot1}
\end{figure}

So far, the fields have not be explicitly specified in the metasurface described by~\eqref{eq:birefsystem}. Since the metasurface can perform arbitrary transformations under the reservation of reciprocity, it may for instance by used for polarization rotation, which will turn to be a most instructive example here. Consider the reflectionless metasurface, depicted in Fig.~\ref{fig:Faraday_rot1}, which transforms the polarization of a normally incident plane wave. The fields corresponding to this transformation are
\begin{subequations}
\label{eq:polrotinc}
\begin{align}
\ve{E}_{\text{i}}(x,y)&=\ve{\hat{x}}\cos(\pi/8)+\ve{\hat{y}}\sin(\pi/8),\\
\ve{H}_{\text{i}}(x,y)&=\frac{1}{\eta_0}\left[-\ve{\hat{x}}\sin(\pi/8)+\ve{\hat{y}}\cos(\pi/8)\right],
\end{align}
\end{subequations}
\begin{subequations}
\begin{align}
\ve{E}_{\text{r}}(x,y)&=0,\\
\ve{H}_{\text{r}}(x,y)&=0,
\end{align}
\end{subequations}
and
\begin{subequations}
\label{eq:polrottrans}
\begin{align}
\ve{E}_{\text{t}}(x,y)&=\ve{\hat{x}}\cos(11\pi/24)+\ve{\hat{y}}\sin(11\pi/24),\\
\ve{H}_{\text{t}}(x,y)&=\frac{1}{\eta_0}\left[-\ve{\hat{x}}\sin(11\pi/24)+\ve{\hat{y}}\cos(11\pi/24)\right].
\end{align}
\end{subequations}
Inserting these fields into~\eqref{eq:field_diff} and~\eqref{eq:field_av}, and substituting the result in~\eqref{eq:chi_diag} yields the susceptibilities
\begin{subequations}
\label{eq:chi_reciprocal_rot}
\begin{align}
\chi_{\text{ee}}^{xx}&=\chi_{\text{mm}}^{yy}=-\frac{1.5048}{k_0}j,\label{eq:chi_reciprocal_rot_a}\\
\chi_{\text{ee}}^{yy}&=\chi_{\text{mm}}^{xx}=\frac{0.88063}{k_0}j.\label{eq:chi_reciprocal_rot_b}
\end{align}
\end{subequations}
Note that in this example\footnote{Incidently, the equality between the electric and magnetic susceptibilities results from the specification of zero reflection in addition to normal incidence. The reader may easily verify that in the presence of reflection, the equalities do not hold.}, the aforementioned double transformation reduces to a single transformation, ${\cal N}(4)=1=4/4$, because the specified fields possess both $x$ and $y$ polarizations. The susceptibilities do not depend on position since the specified transformation, being purely normal, only rotates the polarization angle and does not affect the direction of wave propagation.

The negative and positive imaginary natures of $\chi_{\text{ee}}^{xx}=\chi_{\text{mm}}^{yy}$ and $\chi_{\text{ee}}^{yy}=\chi_{\text{mm}}^{xx}$ in~\eqref{eq:chi_reciprocal_rot} correspond to absorption and gain, respectively. These features may be understood by noting, with the help of Fig.~\ref{fig:Faraday_rot1}, that polarization rotation is accomplished here by attenuation and amplification of $(E_{x,\text{i}},H_{y,\text{i}})$ and  $(E_{y,\text{i}},H_{x,\text{i}})$, respectively. Moreover, this metasurface can rotate the polarization \emph{only}  by the angle $\pi/3$ when the incident wave is polarized at a $\pi/8$ angle\footnote{If, for instance, the incident was polarized along $x$ only, then only the susceptibilities in~\eqref{eq:chi_reciprocal_rot_a} would be excited and the resulting transmitted field would still be polarized along $x$, just with a reduced amplitude with respect to that of the incident wave due to the loss induced by these susceptibilities.}. This example certainly represents an awkward approach to rotate the field polarization!

A more reasonable approach is to consider a \emph{gyrotropic} metasurface, where the only nonzero susceptibilities are $\chi_{\text{ee}}^{xy}$, $\chi_{\text{ee}}^{yx}$, $\chi_{\text{mm}}^{xy}$ and $\chi_{\text{mm}}^{yx}$. This corresponds to a different quadruplet of tensor parameters than in~\eqref{eq:birefsystem}, which illustrates the aforementioned multiplicity of possible parameter set selection. With these susceptibilities, the system~\eqref{eq:InvProbMatrix} yields the following relations:
\begin{subequations}
\label{eq:chi_off_diag}
\begin{align}
\chi_{\text{ee}}^{xy}&=\frac{j\Delta H_{y}}{\omega\epsilon_0  E_{y,\text{av}}},\\
\chi_{\text{ee}}^{yx}&=\frac{-j\Delta H_{x}}{\omega\epsilon_0  E_{x,\text{av}}},\\
\chi_{\text{mm}}^{xy}&=\frac{-j\Delta E_{y}}{\omega\mu_0  H_{y,\text{av}}},\\
\chi_{\text{mm}}^{yx}&=\frac{j\Delta E_{x}}{\omega\mu_0  H_{x,\text{av}}},
\end{align}
\end{subequations}
which, upon substitution of the fields in~\eqref{eq:polrotinc} to~\eqref{eq:polrottrans}, become
\begin{subequations}
\label{eq:pol_rot_imag_chi}
\begin{align}
\chi_{\text{ee}}^{xy}&=\chi_{\text{mm}}^{xy}=-\frac{1.1547}{k_0}j,\\
\chi_{\text{ee}}^{yx}&=\chi_{\text{mm}}^{yx}=\frac{1.1547}{k_0}j.
\end{align}
\end{subequations}
Contrary to the susceptibilities in~\eqref{eq:chi_reciprocal_rot}, those in~\eqref{eq:pol_rot_imag_chi} perform the specified $\pi/3$ polarization rotation \emph{irrespectively} of the initial polarization of the incident wave, due to the gyrotropic nature of the metasurface. It appears that these susceptibilities violate the reciprocity conditions in~\eqref{eq:reciprocity}, and the metasurface is thus \emph{nonreciprocal}, which is a necessary  condition for polarization rotation with this choice of susceptibilities. Thus, the metasurface is a Faraday rotation surface, whose direction of polarization rotation is independent of the direction of wave propagation~\cite{Kodera_APL_07_2011,Sounas_APL_01_2011}. However, contrary to conventional Faraday rotators~\cite{kong1986electromagnetic}, this metasurface is also reflectionless due to the presence of both electric and magnetic gyrotropic susceptibility components (Huygens matching). The positive and negative imaginary susceptibilities indicate that the metasurface is simultaneously active and lossy, respectively. It is this combination of gain and loss that allows perfect rotation in this lossless design. This design is naturally appropriate if Faraday rotation is required. However, it is not optimal in applications not requiring non-reciprocity, i.e. reciprocal gyrotropy, where the required loss and gain would clearly represent a drawback.

Reciprocal gyrotropy may be achieved using bianisotropic chirality, i.e which involves the parameter set $\chi_{\text{em}}^{xx}, \chi_{\text{em}}^{yy}, \chi_{\text{me}}^{xx}$ and $\chi_{\text{me}}^{yy}$. Following the same synthesis procedure as before, we find
\begin{subequations}
\label{eq:pol_rot_chiral}
\begin{align}
\chi_{\text{em}}^{xx}&=\chi_{\text{em}}^{yy}=-\frac{2}{\sqrt{3}k_0}j,\\
\chi_{\text{me}}^{xx}&=\chi_{\text{me}}^{yy}=\frac{2}{\sqrt{3}k_0}j.
\end{align}
\end{subequations}
The corresponding metasurface is readily verified to be reciprocal, passive and lossless, since the susceptibilities~\eqref{eq:pol_rot_chiral} satisfy the conditions~\eqref{eq:lossless}. So, if the purpose of the metasurface is to simply perform polarization rotation in a given direction, without specification for the opposite direction, this design is the most appropriate of the three discussed, as it is purely passive, lossless and working for all incident polarizations.

Note that the metasurfaces~\eqref{eq:pol_rot_imag_chi} and~\eqref{eq:pol_rot_chiral} both correspond to ${\cal N}(4)=1=4/4$.

\subsection{More-Than-Four-Parameter Transformation}
\label{sec:MultiTrans}

In the previous section, we have seen how the system~\eqref{eq:InvProbMatrix} can be solved by reducing the number of susceptibilities to $P=4$ parameters so as to match the number of GSTCs equations, and seen some of the resulting single-transformation (${\cal N}=1$, e.g. monoisotropic structure) or double-transformation (${\cal N}=2$, e.g. birefringence) metasurface possibilities.

However, as mentioned in Sec.~\ref{sec:GenConc}, the general system of equations~\eqref{eq:InvProbMatrix}, given its 16 degrees of freedom (16 susceptibility components), corresponds to a metasurface with the potential capability to perform \emph{more transformations} than a metasurface with 4 parameters, or generally less than 16 parameters, ${\cal N}(16)>{\cal N}(P<16)$. In what follows, we will see how the system~\eqref{eq:InvProbMatrix} can be solved for several \emph{independent} transformations, which includes the possibility of differently processing waves incident from either sides. To accommodate for the additional degrees of freedom, a total of 4 wave transformations are considered, instead of only one as done in Sec.~\ref{sec:singletrans}, so that~\eqref{eq:InvProbMatrix} becomes a full-rank system. The corresponding equations related to the system~\eqref{eq:InvProbMatrix} may then be written in the compact form
\begin{equation}
\begin{split}
\label{eq:fullsystem}
&
\begin{pmatrix}
\Delta H_{y1} & \Delta H_{y2} & \Delta H_{y3} & \Delta H_{y4} \\
\Delta H_{x1} & \Delta H_{x2} & \Delta H_{x3} & \Delta H_{x4} \\
\Delta E_{y1} & \Delta E_{y2} & \Delta E_{y3} & \Delta E_{y4} \\
\Delta E_{x1} & \Delta E_{x2} & \Delta E_{x3} & \Delta E_{x4}
\end{pmatrix}=\\
&\qquad\qquad
\begin{pmatrix}
\widetilde{\chi}_\text{ee}^{xx} & \widetilde{\chi}_\text{ee}^{xy} & \widetilde{\chi}_\text{em}^{xx} & \widetilde{\chi}_\text{em}^{xy}\\
\widetilde{\chi}_\text{ee}^{yx} & \widetilde{\chi}_\text{ee}^{yy} & \widetilde{\chi}_\text{em}^{yx} & \widetilde{\chi}_\text{em}^{yy}\\
\widetilde{\chi}_\text{me}^{xx} & \widetilde{\chi}_\text{me}^{xy} & \widetilde{\chi}_\text{mm}^{xx} & \widetilde{\chi}_\text{mm}^{xy}\\
\widetilde{\chi}_\text{me}^{yx} & \widetilde{\chi}_\text{me}^{yy} & \widetilde{\chi}_\text{mm}^{yx} & \widetilde{\chi}_\text{mm}^{yy}
\end{pmatrix}\\
&\qquad\qquad\qquad
\cdot
\begin{pmatrix}
E_{x1,\text{av}} & E_{x2,\text{av}} & E_{x3,\text{av}} & E_{x4,\text{av}} \\
E_{y1,\text{av}} & E_{y2,\text{av}} & E_{y3,\text{av}} & E_{y4,\text{av}} \\
H_{x1,\text{av}} & H_{x2,\text{av}} & H_{x3,\text{av}} & H_{x4,\text{av}} \\
H_{y1,\text{av}} & H_{y2,\text{av}} & H_{y3,\text{av}} & H_{y4,\text{av}}
\end{pmatrix},
\end{split}
\end{equation}
where the subscripts 1, 2, 3 and 4 indicate the electromagnetic fields corresponding to 4 distinct and \emph{independent} sets of waves\footnote{It is also possible to solve a system of equations that contains less than these 16 susceptibility components. In that case, less than 4 wave transformations should be specified so that the system remains fully determined. For instance, two independent wave transformations (possessing both $x$ and $y$ polarizations) could be solved with 8 susceptibilities. Similarly, 3 wave transformations could be solved with 12 susceptibilities.}. The susceptibilities can be obtained by matrix inversion conjointly with the normalization~\eqref{eq:conv}. The resulting susceptibilities will, in general, be all different from each other. This means that the corresponding metasurface is both active/lossy and nonreciprocal.

Consider for example a metasurface with $P=8$ parameters. In such a case, the system~\eqref{eq:InvProbMatrix} is under-determined, since it features 4 equations in 8 unknowns. This suggests the possibility to specify more than 1 transformations, ${\cal N}>1$. Let us thus consider for instance a monoanisotropic (8-parameter) metasurface, and let us see whether such a metasurface can indeed perform 2 transformations. The corresponding system for 2 transformation reads
\begin{equation}
\label{eq:T2transfo}
\begin{split}
&
\begin{pmatrix}
\Delta H_{y1} & \Delta H_{y2} \\
\Delta H_{x1} & \Delta H_{x2} \\
\Delta E_{y1} & \Delta E_{y2} \\
\Delta E_{x1} & \Delta E_{x2}
\end{pmatrix}=\\
&\quad
\begin{pmatrix}
\widetilde{\chi}_\text{ee}^{xx} & \widetilde{\chi}_\text{ee}^{xy} &
0 & 0\\
\widetilde{\chi}_\text{ee}^{yx} & \widetilde{\chi}_\text{ee}^{yy} &
0 & 0\\
0 & 0
 & \widetilde{\chi}_\text{mm}^{xx} & \widetilde{\chi}_\text{mm}^{xy}\\
0 & 0 &
\widetilde{\chi}_\text{mm}^{yx} & \widetilde{\chi}_\text{mm}^{yy}
\end{pmatrix}\cdot
\begin{pmatrix}
E_{x1,\text{av}} & E_{x2,\text{av}}  \\
E_{y1,\text{av}} & E_{y2,\text{av}}  \\
H_{x1,\text{av}} & H_{x2,\text{av}}  \\
H_{y1,\text{av}} & H_{y2,\text{av}}
\end{pmatrix}.
\end{split}
\end{equation}
This system~\eqref{eq:T2transfo}, being full-rank, automatically admits a solution for the 8~susceptibilities, i.e. ${\cal N}=2$. The only question is whether this solution complies with practical design constraints. For instance, the electric and magnetic susceptibility submatrices are non-diagonal, and may therefore violate the reciprocity condition~\eqref{eq:reciprocity}. If nonreciprocity is undesirable or unrealizable in a practical situation, then one would have to try another set of 8 parameters.

If this 8-parameter metasurface performs only 2 transformations, then one may wonder what is the difference with the 4-parameter birefringent metasurface in~\eqref{eq:birefsystem} which can also provide 2 transformations with just 4 parameters. The difference is that the 2-transformation property of the metasurface in~\eqref{eq:birefsystem} is restricted to the case where the fields of the two transformations are orthogonally polarized\footnote{For instance, if the fields of the first transformation are only $x$-polarized, while the fields of the second transformation are only $y$-polarized.}, whereas the 2-transformation property of the metasurface in~\eqref{eq:T2transfo} is completely general.

\begin{figure}[htbp]
\begin{center}
\subfloat[]{\label{fig:doubletransfo1}
\includegraphics[width=0.5\columnwidth]{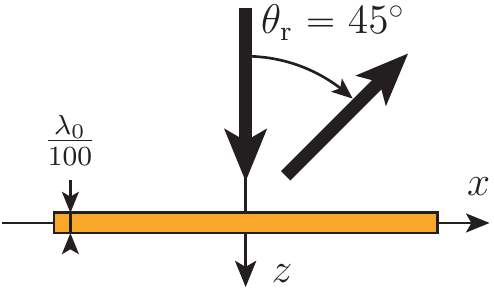}
}
\subfloat[]{\label{fig:doubletransfo2}
\includegraphics[width=0.5\columnwidth]{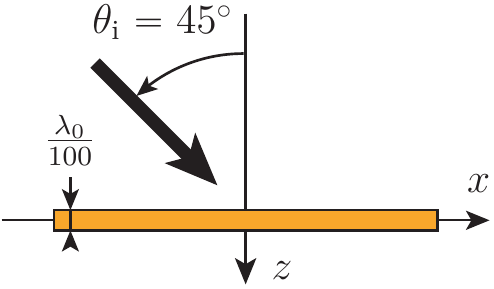}
}
\caption{Example of double-transformation metasurface. (a)~First transformation [corresponding to the subscript~1 in~\eqref{eq:T2transfo}]: the normally incident plane wave is fully reflected at a $45^\circ$ angle. (b)~Second transformation [corresponding to the subscript~2 in~\eqref{eq:T2transfo}]: the obliquely incident plane wave is fully absorbed.}
\label{fig:doubletransfo}
\end{center}
\end{figure}

As an illustration of the latter metasurface, consider the two transformations depicted in Figs.~\ref{fig:doubletransfo}. The first transformation, shown in Fig.~\ref{fig:doubletransfo1}, consists in reflecting at $45^\circ$ a normally incident plane wave. The second transformation, shown in Fig.~\ref{fig:doubletransfo2}, consists in fully absorbing an incident wave impinging on the metasurface under $45^\circ$. In both cases, the transmitted field is specified to be zero for the first and second transformations. The transverse components of the electric fields for the two transformations are, at $z=0$, given by
\begin{subequations}
\label{eq:Especdbltrs}
\begin{equation}\label{eq:Especdbltrs_1}
\ve{E}_\text{i,1} = \frac{\sqrt{2}}{2}(\ve{\hat{x}}+\ve{\hat{y}}),\;
\ve{E}_\text{r,1} = \frac{\sqrt{2}}{2}(-\cos{\theta_\text{r}}\ve{\hat{x}}+\ve{\hat{y}})e^{-jk_xx},
\end{equation}
\begin{equation}\label{eq:Especdbltrs_2}
\ve{E}_\text{i,2} = \frac{\sqrt{2}}{2}(\cos{\theta_\text{i}}\ve{\hat{x}}+\ve{\hat{y}})e^{-jk_xx},
\end{equation}
\end{subequations}
respectively.

The synthesis is then performed by inserting the electric fields~\eqref{eq:Especdbltrs}, and the corresponding magnetic fields, into~\eqref{eq:T2transfo}. The susceptibilities are then straightforwardly obtained by matrix inversion in~\eqref{eq:T2transfo}. For the sake of conciseness, we do not give them here, but we point out that they include nonreciprocity, loss and gain, and complex spatial variations.

\begin{figure}[h!]
\begin{center}
\subfloat[]{\label{fig:COMSOLdlb1}
\includegraphics[width=1\columnwidth]{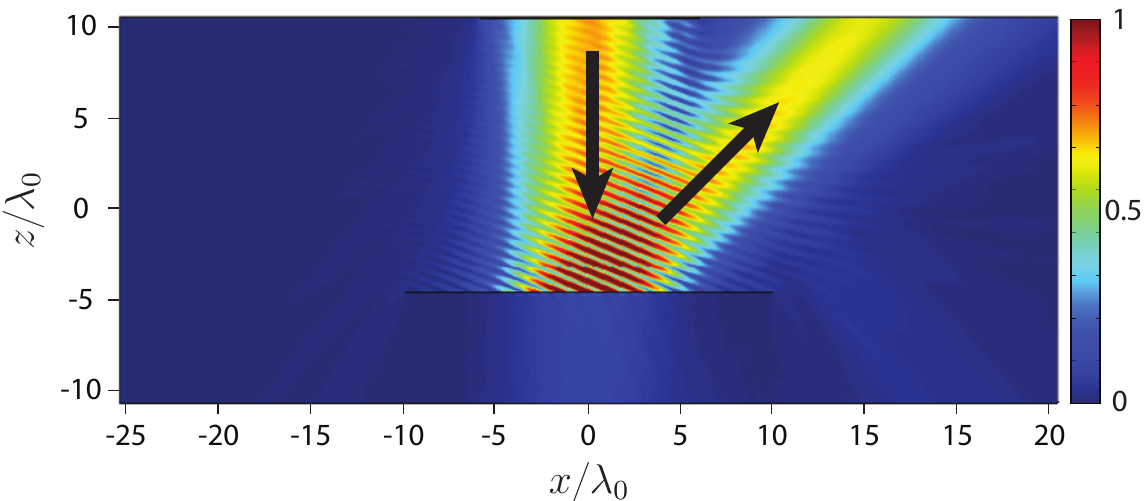}
}\\
\subfloat[]{\label{fig:COMSOLdlb2}
\includegraphics[width=1\columnwidth]{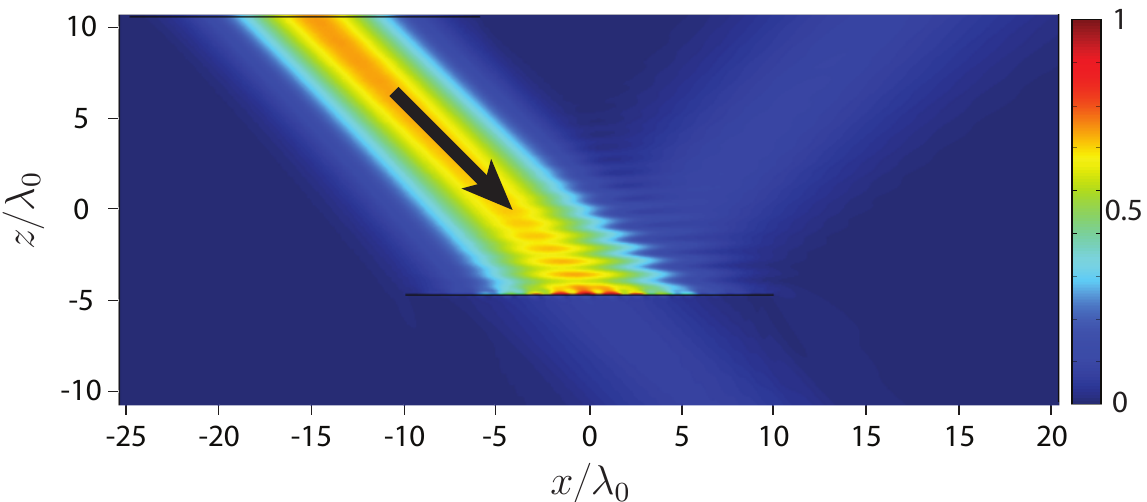}
}
\caption{8-parameter metasurface simulations. COMSOL simulated normalized absolute value of the total electric field corresponding to: (a) the transformation in Fig.~\ref{fig:doubletransfo1}, and (b) the transformation in Fig.~\ref{fig:doubletransfo2}.}
\label{fig:COMSOLdlb}
\end{center}
\end{figure}

This double-transformation response is verified by full-wave simulation and the resulting simulations are plotted in Figs.~\ref{fig:COMSOLdlb}. The two simulations in this figure have been realized in the commercial FEM software COMSOL, where the metasurface is implemented as a thin material slab of thickness $d = \lambda_0/100$\footnote{The synthesis technique yields the susceptibilities for an ideal zero-thickness metasurface. However, the metasurface sheet may be \emph{approximated} by an electrically thin slab of thickness $d$ ($d\ll\lambda$) with volume susceptibility corresponding to a diluted version of the surface susceptibility, i.e. $\chi_\text{vol}=\chi/d$~\cite{achouri2014general}.}. The simulation corresponding to the transformation of Fig.~\ref{fig:doubletransfo1} is shown in Fig.~\ref{fig:COMSOLdlb1}, while the simulation corresponding to the transformation of Fig.~\ref{fig:doubletransfo2} is shown in Fig.~\ref{fig:COMSOLdlb2}. The simulated results are in agreement with the specification [Eq.~\eqref{eq:Especdbltrs}], except for some scattering due to the non-zero thickness of the full-wave slab approximation.

The example just presented, where both the transformations~1 and~2 include all the components of the fields, corresponds to ${\cal N}(8)=2=8/4$, i.e. ${\cal N }(P)=P/4$. However, in the same manner as the birefringent metasurface of~\eqref{eq:chi_diag}, featuring ${\cal N}(4)=2>4/4$, i.e. specifically ${\cal N }(P)=P/2$, the metasurface in~\eqref{eq:T2transfo} may lead to ${\cal N }(P)>P/4$. This depends essentially on whether the specified transformations are composed of fields that are either: a) only $x$- \emph{or} $y$-polarized, or b) both $x$- \emph{and} $y$-polarized. The two transformations given by the fields~\eqref{eq:Especdbltrs} are both $x$- and $y$-polarized, which thus limits the number of transformations to ${\cal N }(P)=P/4$. If the transformation given by~\eqref{eq:Especdbltrs_2} was specified such that $E_\text{iy,2}=0$ (i.e. no polarization along $y$), then this would release degrees of freedom, and hence allow a triple transformation, i.e. ${\cal N}(8)=3>8/4$. In addition, if the first transformation, given by~\eqref{eq:Especdbltrs_1}, also had transverse components of the electric field polarized only along $x$ or $y$, then we could achieve ${\cal N}(8)=4>8/4$ transformations. These considerations illustrate the necessity to perform educated selections in the metasurface synthesis procedure, as announced in Sec.~\eqref{sec:GenConc}.

\subsection{Metasurface with Nonzero Normal Polarizations}
\label{sec:PzMz}

So far, we have discarded the possibility of normal polarizations by enforcing $P_z = M_z = 0$ in~\eqref{eq:BC}. This is not only synthesis-wise convenient, since this suppresses the spatial derivatives in~\eqref{eq:BC}, but also typically justified by the fact that any electromagnetic field can be produced by purely tangential surface currents/polarizations according to the Huygens theorem. It was accordingly claimed in~\cite{8072896} that these normal polarizations, and corresponding susceptibility components, do not bring about any additional degrees of freedom and can thus be completely ignored. It turns out that this claim is generally not true: in fact $P_z$ and $M_z$ provide extra degrees of freedom that allow a metasurface to perform a larger number of distinct operations for \emph{different incident field configurations} and \emph{at different times}.

The Huygens theorem \emph{exclusively} applies to a \emph{single} (arbitrarily complex) combination of incident, reflected and transmitted waves. This means that any metasurface, possibly involving normal polarizations, that performs the specified operation for such a single combination of fields can be reduced to an equivalent metasurface with purely transverse polarizations. However, the Huygens theorem does not apply to case of waves impinging on the metasurface at \emph{different times}. Indeed, it is in this case impossible to superimpose the different incident waves to form a total incident field since they are not simultaneously illuminating the metasurface. Consequently, a purely tangential description of the metasurface is incomplete, and normal polarizations thus become necessary to perform the synthesis.

In fact, the presence of these normal susceptibility components greatly increases the number of degrees of freedom since the susceptibility tensors are now $3\times 3$ matrices, instead of $2\times 2$ as in~\eqref{eq:InvProbMatrix}. This means that, for the 4 relevant GSTCs equations, we have now access to 36~unknown susceptibilities, instead of only 16, which increases the potential number of electromagnetic transformations from 4 to 9, provided that these transformations include fields that are independent from each other.

The synthesis of metasurfaces with nonzero normal polarization densities may be performed following similar procedures as those already discussed. As before, one needs to balance the number of unknown susceptibilities to the number of available equations provided by the GSTCs. Depending on the specifications, this may become difficult since many transformations may be required to obtain a full-rank system. Additionally, if the specified transformations involve changing the direction of wave propagation, then the system~\eqref{eq:BC} becomes a coupled system of partial differential equations in terms of the susceptibilities since the latter would now depend on the position.  This generally prevents the derivation of closed-form solutions of the susceptibilities, which should rather be obtained numerically. However, we will now provide an example of a synthesis problem, where the susceptibilities are obtainable in closed form.

\begin{figure}[htbp]
\begin{center}
\includegraphics[width=0.55\columnwidth]{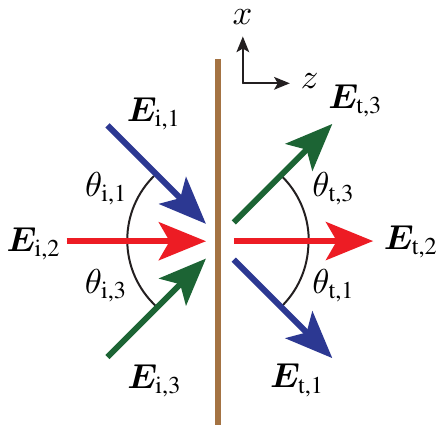}
\caption{Multiple scattering from a uniform bianisotropic reflectionless metasurface.}
\label{Fig:schem}
\end{center}
\end{figure}

More specifically, we discuss the synthesis and analysis of a reciprocal metasurface with controllable angle-dependent scattering~\cite{Gordon2009,DiFalco2011,Radi2015}. To synthesize this metasurface, we consider the three \emph{independent}\footnote{It is essential to understand that these three sets of incident and transmitted waves \emph{cannot} be combined, by superposition, into a single incident and a single transmitted wave because these waves are not necessarily impinging on the metasurface at the same time. This means that the Huygens theorem cannot be used to find purely tangential equivalent surface currents corresponding to these fields.} transformations depicted in Fig.~\ref{Fig:schem}.
Specifying these three transformations allows one to achieve a relatively smooth control of the scattering response of the metasurface for any non-specified incidence angles.

For simplicity, we specify that the metasurface does not change the direction of wave propagation, which implies that it is uniform, i.e. susceptibilities are not functions of position. Moreover, we specify that it is also reflectionless and only affects the transmission phase of  p-polarized incident waves as function of their incidence angle.

To design this metasurface, we consider that it may be composed of a total number of 36 susceptibility components. However, since all the waves interacting with the metasurface are p-polarized, most of these susceptibilities will not be excited by these fields and, thus, will not play a role in the electromagnetic transformations. Accordingly, the only susceptibilities that are excited by the fields are
\begin{subequations}
\label{eq:susc}
\begin{equation}
\te{\chi}_\text{ee}=
\begin{pmatrix}
{\chi}_\text{ee}^{xx} & 0 & {\chi}_\text{ee}^{xz} \\
0 & 0 & 0 \\
{\chi}_\text{ee}^{zx} & 0 & {\chi}_\text{ee}^{zz}
\end{pmatrix},
\quad
\te{\chi}_\text{em}=
\begin{pmatrix}
0 & {\chi}_\text{em}^{xy} & 0\\
0 & 0 & 0\\
0 & {\chi}_\text{em}^{zy} & 0\\
\end{pmatrix},
\end{equation}
\begin{equation}
\te{\chi}_\text{me}=
\begin{pmatrix}
0 & 0 & 0 \\
{\chi}_\text{me}^{yx} & 0 & {\chi}_\text{me}^{yz} \\
0 & 0 & 0
\end{pmatrix},
\quad
\te{\chi}_\text{mm}=
\begin{pmatrix}
0 & 0 & 0\\
0 & {\chi}_\text{mm}^{yy} & 0\\
0 & 0 & 0
\end{pmatrix},
\end{equation}
\end{subequations}
where the susceptibilities not excited have been set to zero for simplicity. In order to satisfy the aforementioned specification of reciprocity, the conditions~\eqref{eq:reciprocity} must be satisfied. This implies that ${\chi}_\text{ee}^{xz} = {\chi}_\text{ee}^{zx}$, ${\chi}_\text{em}^{xy} = -{\chi}_\text{me}^{yx}$ and ${\chi}_\text{em}^{zy} = -{\chi}_\text{me}^{yz}$. As a consequence, the total number of independent susceptibility components in~\eqref{eq:susc} reduces from 9 to 6.

Upon insertion of~\eqref{eq:susc}, the GSTCs in~\eqref{eq:CurlH} and~\eqref{eq:CurlE} become
\begin{subequations}
\label{eq:sys}
\begin{equation}
\Delta H_y = -j\omega\epsilon_0(\chi_\text{ee}^{xx}E_{x,\text{av}} + \chi_\text{ee}^{xz}E_{z,\text{av}}) - jk_0\chi_\text{em}^{xy}H_{y,\text{av}},
\end{equation}
\begin{equation}
\begin{split}
\Delta E_x =& -j\omega\mu_0\chi_\text{mm}^{yy}H_{y,\text{av}}+jk_0(\chi_\text{em}^{xy}E_{x,\text{av}} + \chi_\text{em}^{zy}E_{z,\text{av}})\\
 &-\chi_\text{ee}^{xz} \partial_x E_{x,\text{av}} -\chi_\text{ee}^{zz} \partial_x E_{z,\text{av}} - \eta_0 \chi_\text{em}^{zy} \partial_x H_{y,\text{av}},
\end{split}
\end{equation}
\end{subequations}
where the spatial derivatives only apply to the fields and not to the susceptibilities since the latter are not functions space due to the uniformity of the metasurface.

The system~\eqref{eq:sys} contains 2 equations in 6 unknown susceptibilities and is thus under-determined. In order to solve it, we apply the multiple transformation concept discussed in Sec.~\ref{sec:MultiTrans}, which consists in specifying three independent sets of incident, reflected and transmitted waves. These fields can be simply defined by their respective reflection ($R$)\footnote{Here $R=0$ since the metasurface is reflectionless by specification.} and transmission ($T$) coefficients as well as their incidence angle ($\theta_\text{i}$). In our case, the metasurface exhibits a transmission phase shift, $\phi$, that is function of the incidence angle, i.e. $T = e^{j\phi(\theta_\text{i})}$.

Let us consider, for instance, that the 3 incident plane waves impinge on the metasurface at $\theta_{\text{i},1}=-45^\circ$, $\theta_{\text{i},2}=0^\circ$ and $\theta_{\text{i},3}=+45^\circ$, and are transmitted at $\theta_\text{t}=\theta_\text{i}$ with transmission coefficients $T_1 = e^{-j\alpha}$, $T_2 = 1$ and $T_3 = e^{j\alpha}$, where $\alpha$ is a given phase shift. Solving relations~\eqref{eq:sys} with these specifications yields the following nonzero susceptibilities:
\begin{equation}
\label{eq:suscAPS}
\chi_\text{ee}^{xz} = \chi_\text{ee}^{zx} = \frac{2\sqrt{2}}{k_0}\tan{\left(\frac{\alpha}{2}\right)}.
\end{equation}
It can be easily verified that these susceptibilities satisfy the reciprocity, passivity and losslessness conditions~\eqref{eq:lossless}.

Since the susceptibilities~\eqref{eq:suscAPS} correspond to the only solution of the system~\eqref{eq:sys} for our specifications and since these susceptibilities correspond to the excitation of normal polarization densities, the normal polarizations are indeed useful and provide additional degrees of freedom. This proves the claim in the first paragraph of this section that normal polarizations lead to metasurface functionalities that are unattainable without them.

Now that the metasurface has been synthesized, we analyze its scattering response for all (including non-specified) incidence angles. For this purpose, we substitute the susceptibilities~\eqref{eq:suscAPS} into~\eqref{eq:sys} and consider an incident wave, impinging on the metasurface at an angle $\theta_\text{i}$, being reflected and transmitted with unknown scattering parameters. The system~\eqref{eq:sys} can then be solved to obtain these unknown scattering parameters for any value of $\theta_\text{i}$. In our case, the analysis is simple because the metasurface is uniform, which means that the reflected and transmitted waves obey Snell laws. The resulting angular dependent transmission coefficient is
\begin{equation}
\label{eq:trans2}
T(\theta_\text{i}) = -1 + \frac{2}{1-j\sqrt{2}\sin(\theta_\text{i})\tan{\left(\frac{\alpha}{2}\right)}},
\end{equation}
while the reflection coefficient is $R(\theta_\text{i}) = 0$.

\begin{figure}[ht!]
\begin{center}
\subfloat[]{\label{Fig:abs2}
\includegraphics[width=0.75\columnwidth]{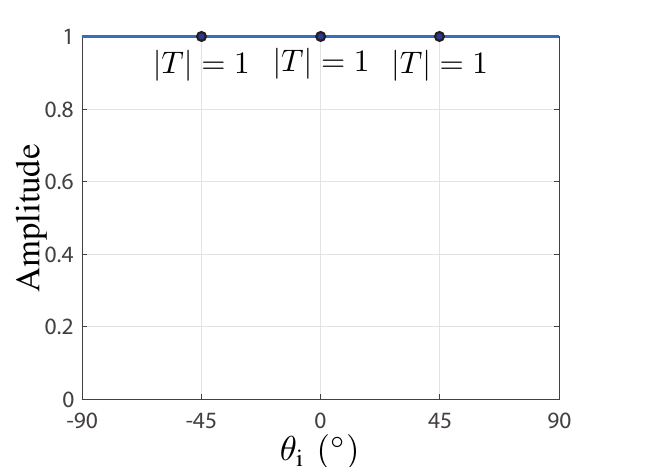}
}\\
\subfloat[]{\label{Fig:phase2}
\includegraphics[width=0.75\columnwidth]{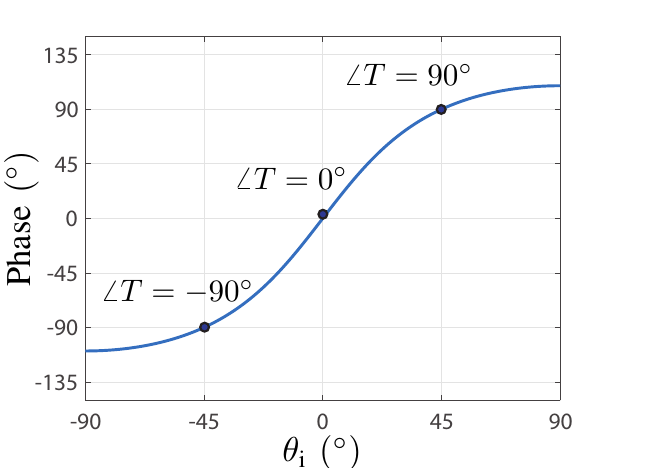}
}
\caption{Transmission amplitude (a) and phase (b) as functions of the incidence angle for a metasurface synthesize for the transmission coefficients $T = \{e^{-j90^\circ};1;e^{j90^\circ}\}$ (and $R=0$) at the respective incidence angles $\theta_\text{i} = \{-45^\circ;0^\circ;+45^\circ\}$.}
\label{Fig:example2}
\end{center}
\end{figure}

In order to illustrate the angular behavior of the transmission coefficient in~\eqref{eq:trans2}, it is plotted in Figs.~\ref{Fig:example2} for a specified phase shift of $\alpha = 90^\circ$. As expected, the transmission amplitude remains unity for all incidence angles while the transmission phase is asymmetric around broadside and covers about a $220^\circ$-phase range.

\subsection{Relations with Scattering Parameters}
\label{sec:Scat}

We have seen how a metasurface can be synthesized so as to obtain its susceptibilities in terms of specified fields. We shall now investigate how the synthesized susceptibilities may be related to the shape of the scattering particles that will constitute the metasurfaces to be realized. Here, we will only present the mathematical expressions that relate the susceptibilities to the scattering particles. The reader is referred to~\cite{Achouri2015c,nl5001746,Pfeiffer2013a,6648706,PhysRevApplied.2.044011,Wong2014360,6905746,6891256,6477089,452013} for more information on the practical realization of these structures.

The conventional method to relate the scattering particle shape to equivalent susceptibilities (or material parameters) is based on homogenization techniques. In the case of metamaterials, these techniques may be used to relate homogenized material parameters to the scattering parameters of the scatterers. From a general perspective, a single isolated scatterer is not sufficient to describe an homogenized medium. Instead, we shall rather consider a periodic array of scatterers, which takes into account the interactions and coupling between adjacent scatterers hence leading to a more accurate description of a ``medium'' compared to a single scatterer. The susceptibilities, which describe the macroscopic responses of a medium, are thus naturally well-suited to describe the homogenized material parameters of metasurfaces. It follows that the equivalent susceptibilities of a scattering particle may be related to the corresponding scattering parameters, conventionally obtained via full-wave simulations, of a periodic array made of an infinite repetition of that scattering particle~\cite{GrbicLightBending,asadchy2011simulation,asadchy2014determining,Pfeiffer2013a}.

Because the periodic array of scatterers is uniform with subwavelength periodicity, the scattered fields obey Snell laws. More specifically, if the incident wave propagates normally with respect to the array, then the reflected and transmitted waves also propagate normally. In most cases, the periodic array of scattering particles is excited with normally propagating waves. This allows one to obtain the 16 \emph{tangential} susceptibility components in~\eqref{eq:fullsystem}. However, it does not provide any information about the normal susceptibility components of the scattering particles. This is because, in the case of normally propagating waves, the normal susceptibilities do not induce any discontinuity of the fields, as explained in Sec.~\ref{sec:GenConc}. Nevertheless, this method allows one to match the tangential susceptibilities of the scattering particle to the susceptibilities found from the metasurface synthesis procedure and that precisely yields the ideal tangential susceptibility components.

It is clear that the scattering particles may, in addition to their tangential susceptibilities, possess nonzero normal susceptibility components. In that case, the scattering response of the metasurface, when illuminated with obliquely propagating waves, will differ from the expected ideal behavior prescribed in the synthesis. Consequently, the homogenization technique only serve as an initial guess to describe the scattering behavior of the metasurface\footnote{Note that is possible to obtain all 36 susceptibility components of a scattering particle provided that the 4 GSTCs relations are solved for 9 independent sets of incident, reflected and transmitted waves. In practice, such an operation is particularity tedious and is thus generally avoided.}.

We will now derive the explicit expressions relating the tangential susceptibilities to the scattering parameters in the general case of a fully bianisotropic uniform metasurface surrounded by different media and excited by normally incident plane waves. Let us first write the system~\eqref{eq:fullsystem} in the following compact form:
\begin{equation}
\label{eq:reducedSys}
 \te{\Delta} = \widetilde{\te{\chi}}\cdot \te{A}_v,
\end{equation}
where the matrices $\te{\Delta}$, $\widetilde{\te{\chi}}$ and $\te{A}_v$ correspond to the field differences, the normalized susceptibilities and the field averages, respectively.

\begin{figure}[ht]
\centering
\includegraphics[width=1\linewidth]{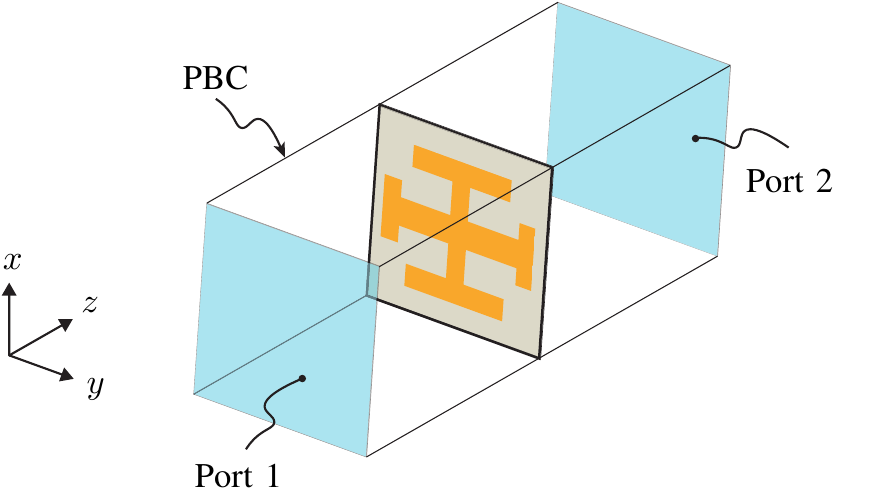}
\caption{Full-wave simulation setup for the scattering parameter technique leading to the metasurface physical structure from the metasurface model based on~\eqref{eq:reducedSys}. The unit cell is surrounded by periodic boundary conditions (PBC) and excited from port~1 and~2.}
\label{fig:UnitCellSim}
\end{figure}

In order to obtain the 16 tangential susceptibility components in~\eqref{eq:fullsystem}, we will now define four transformations by specifying the fields on both sides of the metasurface. Let us consider that the metasurface is illuminated from the left with an $x$-polarized normally incident plane wave. The corresponding incident, reflected and transmitted electric fields read
\begin{equation}
\label{eq:xPol}
\ve{E}_{\text{i}}=\ve{\hat{x}},
\quad
\ve{E}_{\text{r}}=S_{11}^{xx}\ve{\hat{x}} + S_{11}^{yx}\ve{\hat{y}},
\quad
\ve{E}_{\text{t}}=S_{21}^{xx}\ve{\hat{x}} + S_{21}^{yx}\ve{\hat{y}},
\end{equation}
where the terms $S_{ab}^{uv}$, with $a, b = \{1,2\}$ and $u, v = \{x,y\}$, are the scattering parameters with ports 1 and 2 corresponding to the left and right sides of the metasurface, respectively, as shown in Fig.~\ref{fig:UnitCellSim}.
The medium of the left of the metasurface has the intrinsic impedance $\eta_1$, while the medium on the right has the intrinsic impedance $\eta_2$. In addition to~\eqref{eq:xPol}, three other cases have to be considered, i.e. $y$-polarized excitation incident from the left (port 1), and $x$- and $y$-polarized excitations incident from the right (port 2). Inserting these fields into~\eqref{eq:fullsystem}, leads, after simplification, to the matrices $\te{\Delta}$ and $\te{A}_v$ given below,
\begin{subequations}
\begin{floatEq}
\begin{equation}
\label{eq:deltaMat}
\te{\Delta}=
\begin{pmatrix}
-\te{N}_2/\eta_1 + \te{N}_2\cdot\te{S}_{11}/\eta_1 + \te{N}_2\cdot\te{S}_{21}/\eta_2 & -\te{N}_2/\eta_2 +\te{N}_2\cdot\te{S}_{12}/\eta_1 + \te{N}_2\cdot\te{S}_{22}/\eta_2 \\
-\te{N}_1\cdot\te{N}_2 - \te{N}_1\cdot\te{N}_2\cdot\te{S}_{11} + \te{N}_1\cdot\te{N}_2\cdot\te{S}_{21} & \te{N}_1\cdot\te{N}_2 - \te{N}_1\cdot\te{N}_2\cdot\te{S}_{12}+ \te{N}_1\cdot\te{N}_2\cdot\te{S}_{22}
\end{pmatrix},
\end{equation}
\begin{equation}
\label{eq:AvMat}
\te{A}_v=\frac{1}{2}
\begin{pmatrix}
\te{I} + \te{S}_{11}+ \te{S}_{21} &
\te{I} + \te{S}_{12}+ \te{S}_{22}
\\
\te{N}_1/\eta_1 - \te{N}_1\cdot\te{S}_{11}/\eta_1 + \te{N}_1\cdot\te{S}_{21}/\eta_2 &
-\te{N}_1/\eta_2 - \te{N}_1\cdot\te{S}_{12}/\eta_1 + \te{N}_1\cdot\te{S}_{22}/\eta_2
\end{pmatrix}.
\end{equation}
\end{floatEq}
\end{subequations}
where the matrices $\te{S}_{ab}$, $\te{I}$, $\te{N}_1$ and $\te{N}_2$ are defined by
\begin{equation}
\begin{split}
&\te{S}_{ab}=
\begin{pmatrix}
S_{ab}^{xx} & S_{ab}^{xy} \\
S_{ab}^{yx} & S_{ab}^{yy}
\end{pmatrix},\qquad
\te{I}=
\begin{pmatrix}
1 & 0 \\
0 & 1
\end{pmatrix},\\
&\te{N}_1=
\begin{pmatrix}
0 & -1 \\
1 & 0
\end{pmatrix},\qquad
\te{N}_2=
\begin{pmatrix}
1 & 0 \\
0 & -1
\end{pmatrix}.
\end{split}
\end{equation}
Now, the procedure to obtain the susceptibilities of a given scattering particle is as follows: firstly, the scattering particle is simulated with periodic boundary conditions and normal excitation. Secondly, the resulting scattering parameters obtained from the simulations are used to define the matrices in~\eqref{eq:deltaMat} and~\eqref{eq:AvMat}. Finally, the susceptibilities corresponding to the particle are obtained by matrix inversion of~\eqref{eq:reducedSys}.

Alternatively, it is possible to obtain the scattering parameters of a normally incident plane being scattered by a uniform metasurface with known susceptibilities. This can be achieved by solving~\eqref{eq:reducedSys} for the scattering parameters. This leads to the following matrix equation:
\begin{equation}
\label{eq:reducedSysInv}
 \te{S} = \te{M}_1^{-1}\cdot\te{M}_2,
\end{equation}
where the scattering parameter matrix, $\te{S}$, is defined as
\begin{equation}\label{eq:Smatrix}
\te{S}=
\begin{pmatrix}
\te{S}_{11} & \te{S}_{12} \\
\te{S}_{21} & \te{S}_{22}
\end{pmatrix},
\end{equation}
and the matrices $\te{M}_1$ and $\te{M}_2$ are obtained from~\eqref{eq:reducedSys},~\eqref{eq:deltaMat} and~\eqref{eq:AvMat} by expressing the scattering parameters in terms of the normalized susceptibility tensors. The resulting matrices $\te{M}_1$ and $\te{M}_2$ are given below.
\begin{subequations}
\begin{floatEq}
\begin{equation}
\label{eq:Mat1}
\te{M}_1=
\begin{pmatrix}
\te{N}_2/\eta_1 - \widetilde{\te{\chi}}_\text{ee}/2 + \widetilde{\te{\chi}}_\text{em}\cdot\te{N}_1/(2\eta_1) & \te{N}_2/\eta_2 - \widetilde{\te{\chi}}_\text{ee}/2 - \widetilde{\te{\chi}}_\text{em}\cdot\te{N}_1/(2\eta_2) \\
-\te{N}_1\cdot\te{N}_2 - \widetilde{\te{\chi}}_\text{me}/2 + \widetilde{\te{\chi}}_\text{mm}\cdot\te{N}_1/(2\eta_1) & \te{N}_1\cdot\te{N}_2 - \widetilde{\te{\chi}}_\text{me}/2 - \widetilde{\te{\chi}}_\text{mm}\cdot\te{N}_1/(2\eta_2)
\end{pmatrix},
\end{equation}
\begin{equation}
\label{eq:Mat2}
\te{M}_2=
\begin{pmatrix}
\widetilde{\te{\chi}}_\text{ee}/2 + \te{N}_2/\eta_1+\widetilde{\te{\chi}}_\text{em}\cdot\te{N}_1/(2 \eta_1) & \widetilde{\te{\chi}}_\text{ee}/2 + \te{N}_2/\eta_2-\widetilde{\te{\chi}}_\text{em}\cdot\te{N}_1/(2 \eta_2) \\
\widetilde{\te{\chi}}_\text{me}/2 + \te{N}_1\cdot\te{N}_2+\widetilde{\te{\chi}}_\text{mm}\cdot\te{N}_1/(2 \eta_1) & \widetilde{\te{\chi}}_\text{me}/2 - \te{N}_1\te{N}_2-\widetilde{\te{\chi}}_\text{mm}\cdot\te{N}_1/(2 \eta_2)
\end{pmatrix}.
\end{equation}
\end{floatEq}
\end{subequations}

Thus, the final metasurface physical structure is obtained by mapping the scattering parameters~\eqref{eq:Smatrix} obtained from the discretized synthesized susceptibilities by~\eqref{eq:reducedSysInv} via~\eqref{eq:Mat1} and~\eqref{eq:Mat2} to those obtained by full-wave simulating metasurface unit cells with tunable parameters, in an approximate periodic environment, as illustrated in Fig.~\ref{fig:UnitCellSim}.

\section{Concepts and Applications}
\label{sec:ConApp}

In the previous section, we have shown several metasurface examples as \emph{illustrations} of the proposed synthesis technique. These examples did not necessarily correspond to practical designs but, in addition to illustrating the proposed synthesis technique, they did set up the stage for the development of useful and practical concepts and applications, which is the object of the present section.

We shall present here 5 of our most recent works representing novel concepts and applications of metasurfaces. In the order of appearance, we present our work on birefringent transformations~\cite{Achouri2015c,Achouri2016e}, bianisotropic refraction~\cite{Lavigne2017}, light emission enhancement~\cite{Chen2016}, remote spatial processing~\cite{Achouri2016c} and nonlinear second-harmonic generation~\cite{achouri2017mathematical}. The reader is also referred to our related works on nonreciprocal nongyrotropic isolators~\cite{Taravati2016}, dielectric metasurfaces for dispersion engineering~\cite{Achouri2016d} and radiation pressure control~\cite{achouri2017metasurface}.

\subsection{Birefringent Operations}

A direct application of the synthesis procedure discussed in Sec.~\ref{sec:Syn}, and more specifically of the susceptibilities in~\eqref{eq:chi_diag}, is the design of birefringent metasurfaces. These susceptibilities are split into two independent sets that allow to individually control the scattering of s- and p-polarized waves. In particular, the manipulation of the respective transmission phases of these orthogonal waves allows several interesting operations.

\begin{figure}[ht!]
\centering
\includegraphics[width=1\columnwidth]{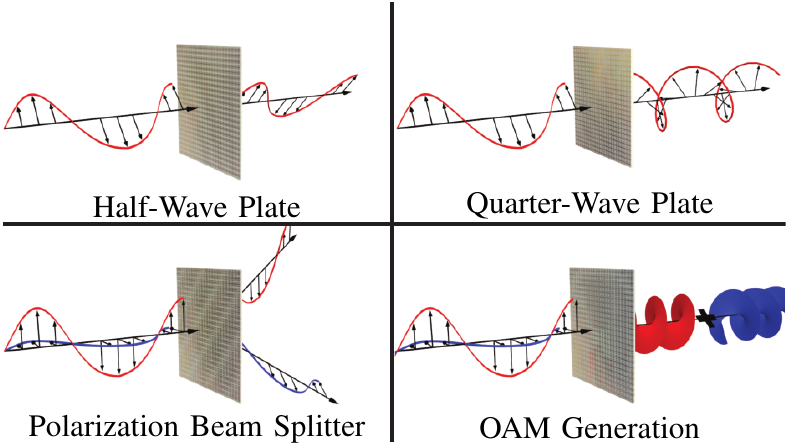}
\caption{Birefringent metasurface transformations presented in~\cite{Achouri2016e}.}
\label{Fig:Blend}
\end{figure}

In~\cite{Achouri2016e}, we have used this approach to realize half-wave plates, which rotate the polarization of linearly polarized waves by $90^\circ$ or invert the handedness of circularly polarized waves, quarter-wave plates, which convert linear polarization into circular polarization, a polarization beam splitter, which spatially separates orthogonally polarized waves, and an orbital angular momentum generator, which generates topological charges that depend on the incident wave polarization. These operations are depicted in Fig.~\ref{Fig:Blend}.

\subsection{``Perfect'' Refraction}

Most refractive operations realized so far with a metasurface have been based on the concept of the generalized law of refraction~\cite{capasso1}, which requires the implementation of a phase gradient structure. However, such structures are plagued by undesired diffraction orders and are thus not fully efficient. It turns out that the fundamental reason for this efficiency limitation is the symmetric nature of simple (early as in~\cite{capasso1}) refractive metasurfaces with respect to the $z$-direction. This can be demonstrated by the following ad absurdum argument.

Let us consider a passive metasurface surrounded by a given reciprocal\footnote{The quasi-totality of the refracting metasurfaces discussed in the literature so far have been reciprocal. The following argument does not hold for the nonreciprocal case, where perfect refraction could in principle be achieved by a symmetric metasurface structure.} medium and denote the two sides of the structure by the indices~1 and~2. Assume that this metasurface \emph{perfectly} refracts (without reflection and spurious diffraction) a wave incident under the angle $\theta_1$ in side~1 to the angle~$\theta_2$ in side~2, and assume, \emph{ad absurdum}, that this metasurface is \emph{symmetric} with respect to its normal. Since it is reciprocally perfectly refracting, it is perfectly matched for both propagation directions, 1~to~2 and 2~to~1. Consider first wave propagation from side~1 to side~2. Due to perfect matching, the wave experiences no reflection and, due to perfect refraction, it is fully transmitted to the angle $\theta_2$ in side~2. Consider now wave propagation in the opposite direction, along the reciprocal (or time-reversed) path. Now, the wave incident in side~2 has different tangential field components than that incident in side~1, assuming $\theta_2\neq\theta_1$, and, therefore, it will see a different impedance, which means that the metasurface is necessarily mismatched in the direction 2~to~1. But this is in contradiction with the assumption of perfect (reciprocal) refraction! Consequently, the symmetric metasurface does not produce perfect refraction. Part of the wave incident from side~2 is reflected back and therefore, by reciprocity, matching also did not actually exist in the direction 1~to~2, so all of the energy of the wave incident under~$\theta_1$ in side~1 cannot completely refract into~$\theta_2$; part of it has to be transmitted to other directions in side~2, which typically represents spurious diffraction orders assuming a periodic-gradient metasurface. These diffraction orders are consistently visible in reported simulations and experiments of symmetric metasurfaces intended to perform refraction.

It was demonstrated in~\cite{7506314,Lavigne2017} that \emph{bianisotropy} was the solution to realize perfect (reciprocal) refraction ($100\%$ power transmission efficiency from~$\theta_1$ to $\theta_2$). In what follows, we summarize the main synthesis steps for such a metasurface.

\begin{figure}[ht]
\centering
\subfloat[]{\label{fig:BianiRef1}
\includegraphics[width=0.5\columnwidth]{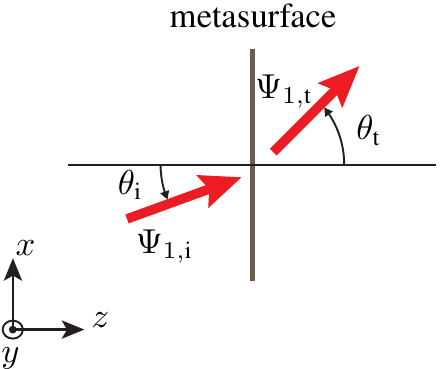}
}
\subfloat[]{\label{fig:BianiRef2}
\includegraphics[width=0.5\columnwidth]{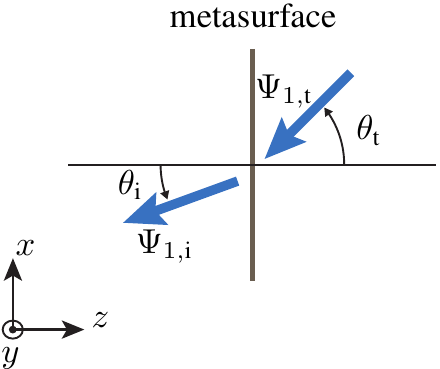}
}
\caption{Representation of the two transformations specified in the system~\eqref{eq:Biani2}. (a)~First transformation, corresponding to the fields in~\eqref{eq:Field1}, (b)~Second transformation, corresponding to the fields in~\eqref{eq:Field2}.}
\label{fig:BianiRef}
\end{figure}

Let us consider the bianisotropic GSTCs relations in~\eqref{eq:InvProbMatrix}. For a refractive metasurface, rotation of polarization is not required and usually undesired. Therefore, the relevant nonzero susceptibility components reduce to the diagonal components of $\te{\chi}_\text{ee}$ and $\te{\chi}_\text{mm}$ and the off-diagonal components of $\te{\chi}_\text{em}$ and $\te{\chi}_\text{me}$. This corresponds to $4\times 2=8$ susceptibility parameters, leading, according to Sec.~\ref{sec:MultiTrans}, to the double-transformation full-rank system
\begin{equation}
\begin{split}
\label{eq:Biani}
&
\begin{pmatrix}
\Delta H_{y1} & \Delta H_{y2}  \\
\Delta H_{x1} & \Delta H_{x2}  \\
\Delta E_{y1} & \Delta E_{y2}  \\
\Delta E_{x1} & \Delta E_{x2}
\end{pmatrix}=\\
&\qquad
\begin{pmatrix}
\widetilde{\chi}_\text{ee}^{xx} & 0 &
0 & \widetilde{\chi}_\text{em}^{xy}\\
0 & \widetilde{\chi}_\text{ee}^{yy} &
\widetilde{\chi}_\text{em}^{yx} & 0\\
0 & \widetilde{\chi}_\text{me}^{xy} &
\widetilde{\chi}_\text{mm}^{xx} & 0\\
\widetilde{\chi}_\text{me}^{yx} & 0 &
0 & \widetilde{\chi}_\text{mm}^{yy}
\end{pmatrix}
\cdot
\begin{pmatrix}
E_{x1,\text{av}} & E_{x2,\text{av}}  \\
E_{y1,\text{av}} & E_{y2,\text{av}}  \\
H_{x1,\text{av}} & H_{x2,\text{av}}  \\
H_{y1,\text{av}} & H_{y2,\text{av}}
\end{pmatrix},
\end{split}
\end{equation}
where we naturally specify the second transformation as the reciprocal of the first one. Assuming that the refraction takes places in the $xz$-plane and that the waves are all p-polarized, the system~\eqref{eq:Biani} reduces to
\begin{equation}
\label{eq:Biani2}
\begin{pmatrix}
\Delta H_{y1} & \Delta H_{y2}  \\
\Delta E_{x1} & \Delta E_{x2}
\end{pmatrix}=
\begin{pmatrix}
\widetilde{\chi}_\text{ee}^{xx} & \widetilde{\chi}_\text{em}^{xy}\\
\widetilde{\chi}_\text{me}^{yx} & \widetilde{\chi}_\text{mm}^{yy}
\end{pmatrix}
\cdot
\begin{pmatrix}
E_{x1,\text{av}} & E_{x2,\text{av}}  \\
H_{y1,\text{av}} & H_{y2,\text{av}}
\end{pmatrix},
\end{equation}
which strictly corresponds to a system that is ${\cal N}(4)=2$ although the initial goal might have been to perform refraction in one propagation direction only. An illustration of the first and second transformations is presented in Figs.~\ref{fig:BianiRef1} and~\ref{fig:BianiRef2}, respectively. Note that the subscripts i and t respectively refer to the incident and transmit sides of the metasurface rather than the incident and transmitted waves.
The electromagnetic fields on the incident and transmit sides of the metasurface, assuming that the media on both sides are vacuum, and that correspond to the first transformation read
\begin{subequations}
\label{eq:Field1}
\begin{align}
E_{x1,\text{i}} &= \frac{k_{z,\text{i}}}{k_0}e^{-jk_{x,\text{i}}x},\qquad E_{x1,\text{t}} = A_\text{t}\frac{k_{z,\text{t}}}{k_0}e^{-jk_{x,\text{t}}x},\\
H_{y1,\text{i}} &= e^{-jk_{x,\text{i}}x}/\eta_0, \qquad H_{y1,\text{t}} =A_\text{t} e^{-jk_{x,\text{t}}x}/\eta_0,
\end{align}
\end{subequations}
where $A_\text{t}$ is the amplitude of the wave on the transmit side. The fields corresponding to the second transformation read
\begin{subequations}
\label{eq:Field2}
\begin{align}
E_{x2,\text{i}} &= -\frac{k_{z,\text{i}}}{k_0}e^{jk_{x,\text{i}}x},\qquad E_{x2,\text{t}} = -A_\text{t}\frac{k_{z,\text{t}}}{k_0}e^{jk_{x,\text{t}}x},\\
H_{y2,\text{i}} &= e^{jk_{x,\text{i}}x}/\eta_0, \quad\qquad H_{y2,\text{t}} =A_\text{t} e^{jk_{x,\text{t}}x}/\eta_0.
\end{align}
\end{subequations}
In order to ensure power conservation between the incident and transmitted waves, the amplitude of the transmitted wave must be $A_\text{t}=\sqrt{k_{z,\text{i}}/k_{z,\text{t}}}=\sqrt{\cos{\theta_\text{i}}/\cos{\theta_\text{t}}}$, as shown in~\cite{Lavigne2017}. Under this condition, the metasurface susceptibilities, obtained by substituting~\eqref{eq:Field1} and~\eqref{eq:Field2} into~\eqref{eq:Biani2} and considering the normalization~\eqref{eq:conv}, read
\begin{subequations}
\label{eq:bianiChi}
\begin{equation}
\chi_\text{ee}^{xx} = \frac{4  \sin ( \alpha x)}{
   \beta \cos ( \alpha x)+  \sqrt{\beta^2-\gamma^2}},
\end{equation}
\begin{equation}
\chi_\text{mm}^{yy} =   \frac{\beta^2-\gamma^2}{4 k_0^2 }    \frac{4   \sin (
   \alpha x)}{ \beta \cos ( \alpha x)+ \sqrt{\beta^2-\gamma^2}},
\end{equation}
\begin{equation}
\chi_\text{em}^{xy} =  -\chi_\text{me}^{yx}= \frac{2j}{k_0}\frac{ \gamma\cos ( \alpha x)}{
   \beta \cos ( \alpha x)+ \sqrt{\beta^2-\gamma^2}},
\end{equation}
\end{subequations}
where $\alpha = k_{x,\text{t}} - k_{x,\text{i}}$, $\beta = k_{z,\text{i}} + k_{z,\text{t}}$ and $\gamma = k_{z,\text{i}} - k_{z,\text{t}}$. It can be easily verified, using~\eqref{eq:lossless}, that the bianisotropic refractive metasurface with the susceptibilities~\eqref{eq:bianiChi} corresponds to a reciprocal, passive and lossless structure, in addition to being immune to reflection and spurious diffraction, and is hence a perfectly refractive metasurface.

\begin{figure}[ht!]
\centering
\subfloat[]{\label{Fig:Demo1}
\includegraphics[width=0.35\columnwidth]{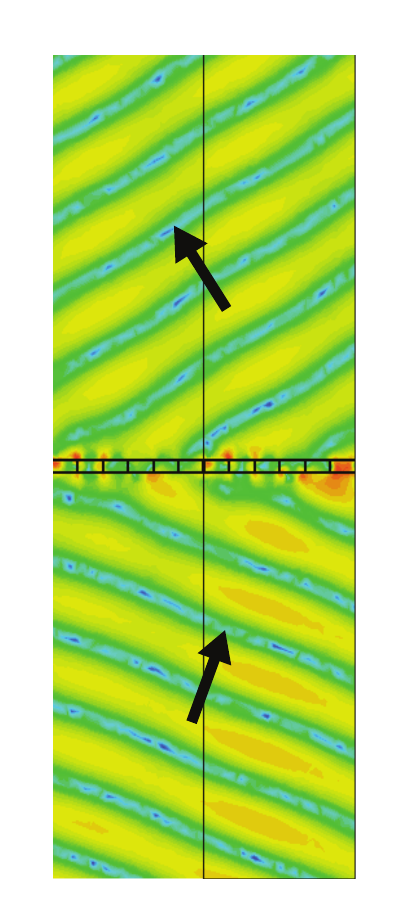}}
\subfloat[]{\label{Fig:Demo2}
\includegraphics[width=0.35\columnwidth]{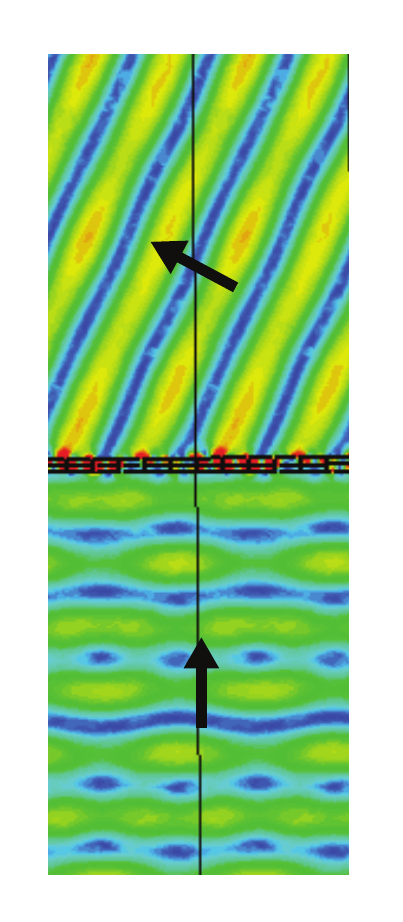}}
\caption{Full-wave simulations showing the performance of two refractive metasurfaces~\cite{Lavigne2017}.}
\label{Fig:Demo}
\end{figure}

To demonstrate the performance of the synthesis method, we have built two bianisotropic refractive metasurfaces~\cite{Lavigne2017}. They respectively transform an incident wave impinging at $\theta_\text{i}=20^\circ$ into a transmitted wave refracted at $\theta_\text{i}=-28^\circ$, and a normally incident wave into a transmitted wave refracted at $\theta_\text{i}=-70^\circ$. The full-wave simulations corresponding to these transformations are respectively plotted in Figs.~\ref{Fig:Demo1} and.~\ref{Fig:Demo2}. The simulated power transmission of these two structures is respectively $86.7\%$ and $83.2\%$. These efficiencies are mostly limited to the inherent dielectric and metallic losses of the scattering particles and, to a lesser extent, to the undesired diffraction orders due to the imperfection of these particles. A corresponding metasurface was demonstrated in~\cite{Lavigne2017} with an efficiency (79~$\%$) that is around 4~$\%$ superior to the theoretical limit of a \emph{lossless} monoanisotropic metasurface, hence unquestionably demonstrating the superiority of the bianisotropic design!

\subsection{Remote Spatial Processing}

Metasurface remote spatial processing, introduced in~\cite{Achouri2016c}, consists in controlling the transmission of a signal beam through a metasurface by remotely sending a control beam, which properly interferes with the signal beam. This interference is thus used to shape the metasurface transmission pattern by varying the phase and/or amplitude of the control beam. 

\begin{figure}[ht]
\centering
\subfloat[]{\includegraphics[width=0.45\linewidth]{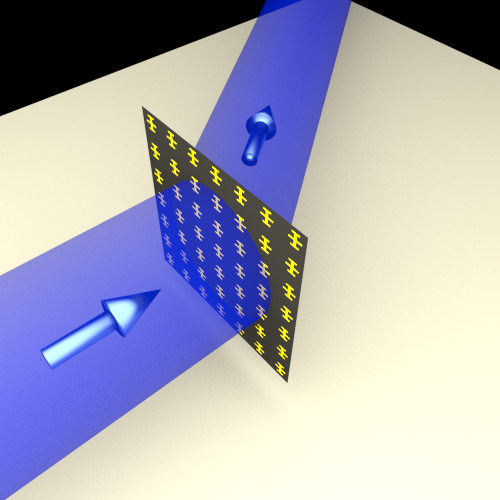}
\label{fig:param_prob3}}
\subfloat[]{\includegraphics[width=0.45\linewidth]{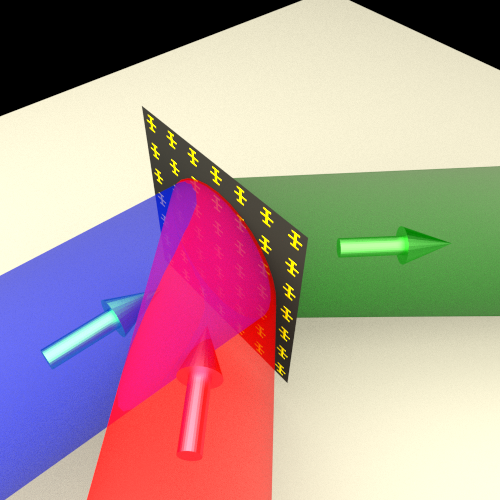}
\label{fig:param_prob4}}
\caption{Example of a remote spatial processing operation. (a)~Signal beam being refracted by the metasurface. (b)~Superposition of signal and control beams interacting with each other, which leads to a different transmitted wave.}\label{fig:param_prob}
\end{figure}

Figure~\ref{fig:param_prob} presents an example of such remote spatial processing. Initially, the signal beam (in blue) in Fig.~\ref{fig:param_prob3} is refracted by the metasurface according to some initial specification. When the control beam (in red) is next added to the signal beam on the metasurface, as in Fig.~\ref{fig:param_prob4}, it changes the overall radiation pattern of the metasurface.

\begin{figure}[ht]
\begin{center}
\includegraphics[width=0.65\columnwidth]{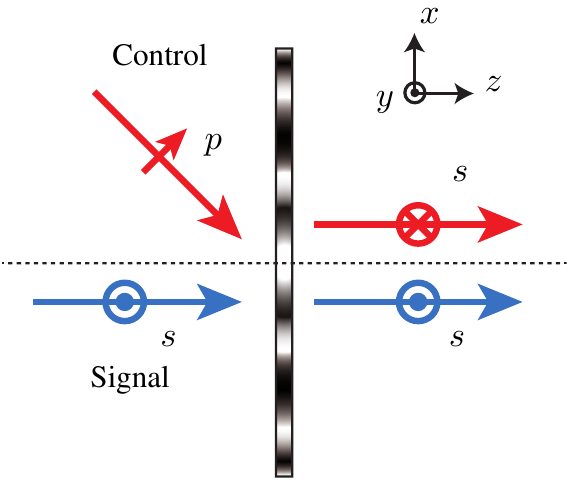}
\caption{Coherent modulator metasurface. The signal and control beams are impinging on the metasurface at different angles to avoid collocation of their source. The amplitude of the transmitted wave depends on the phase difference between the two beams by interference.}\label{Fig:TMconcept}
\end{center}
\end{figure}

We have used this concept to implement remote spatial switch/modulators. The operation principle of such a modulator is presented in Fig.~\ref{Fig:TMconcept}. To avoid the collocation of the control and signal beam sources, the control beam impinges on the metasurface at an angle while the signal beam is normally incident. In order to independently control the transmission of both beams, they must be orthogonally polarized on the incident side of the metasurface. However, they must exhibit the same polarization on the transmit side so as to interfere. In~\cite{Achouri2016c}, we show that such a transformation can only be achieved using a bianisotropic metasurface, which must also be chiral so as to rotate the polarization of the control beam. On the transmit side, the two beams interfere and the corresponding amplitude thus depends of the phase difference between them.

\begin{figure}[ht]
\begin{center}
\includegraphics[width=0.75\columnwidth]{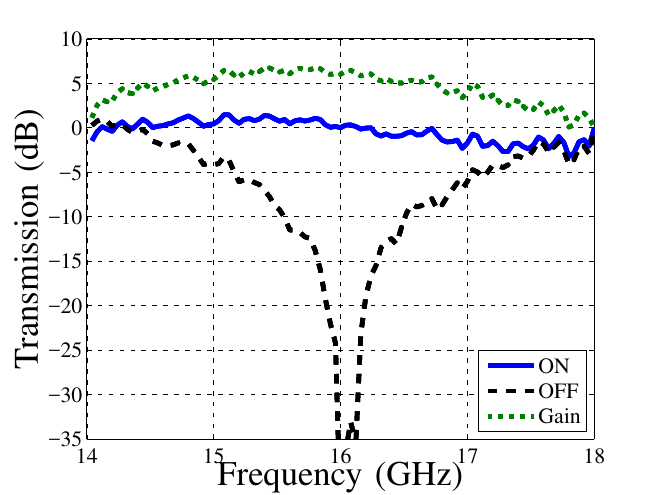}
\caption{Measured transmission coefficients for the metasurface in Fig.~\ref{Fig:TMconcept}. The blue curve is the transmission of the signal beam only, while the black and green curves are the destructive and constructive interferences of the signal and control beams, respectively.}\label{Fig:BWnormal}
\end{center}
\end{figure}

The fabricated metasurface performing the operation depicted in Fig.~\ref{Fig:TMconcept} has been experimentally measured, and the corresponding results are plotted in Fig.~\ref{Fig:BWnormal} for an operating frequency of 16~GHz.

\subsection{Light Emission Enhancement}

In the perspective of enhancing the efficiency of light-emitting diodes (LEDs), we have reported in~\cite{Chen2016} a partially reflecting metasurface cavity (PRMC) increasing the emission of photon sources in layered semiconductor structures, using the susceptibility-GSTC technique presented in this paper. This PRMC simultaneously enhances the light extraction efficiency (LEE), spontaneous emission rate (SER) and far-field directivity of the photon source.

\begin{figure}[ht]
\centering
\subfloat[]{\includegraphics[width=0.5\linewidth]{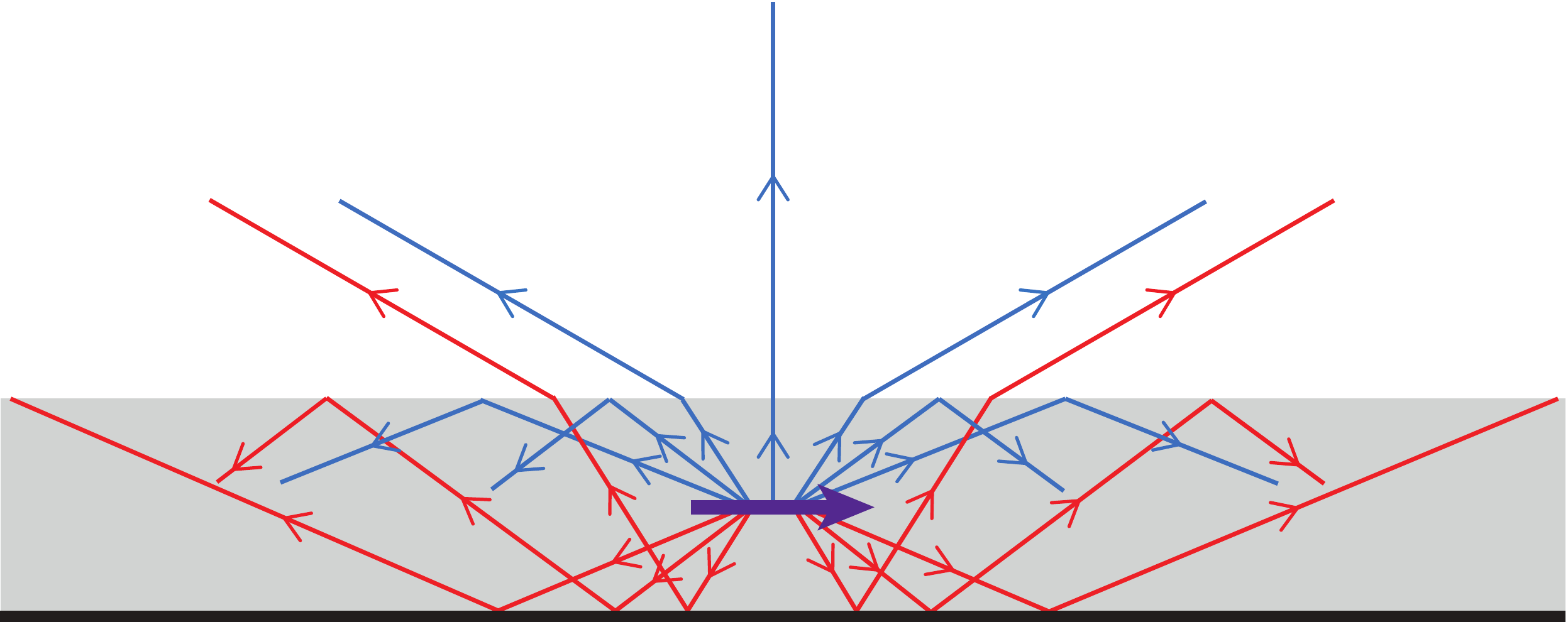}
\label{fig:led1}}
\subfloat[]{\includegraphics[width=0.5\linewidth]{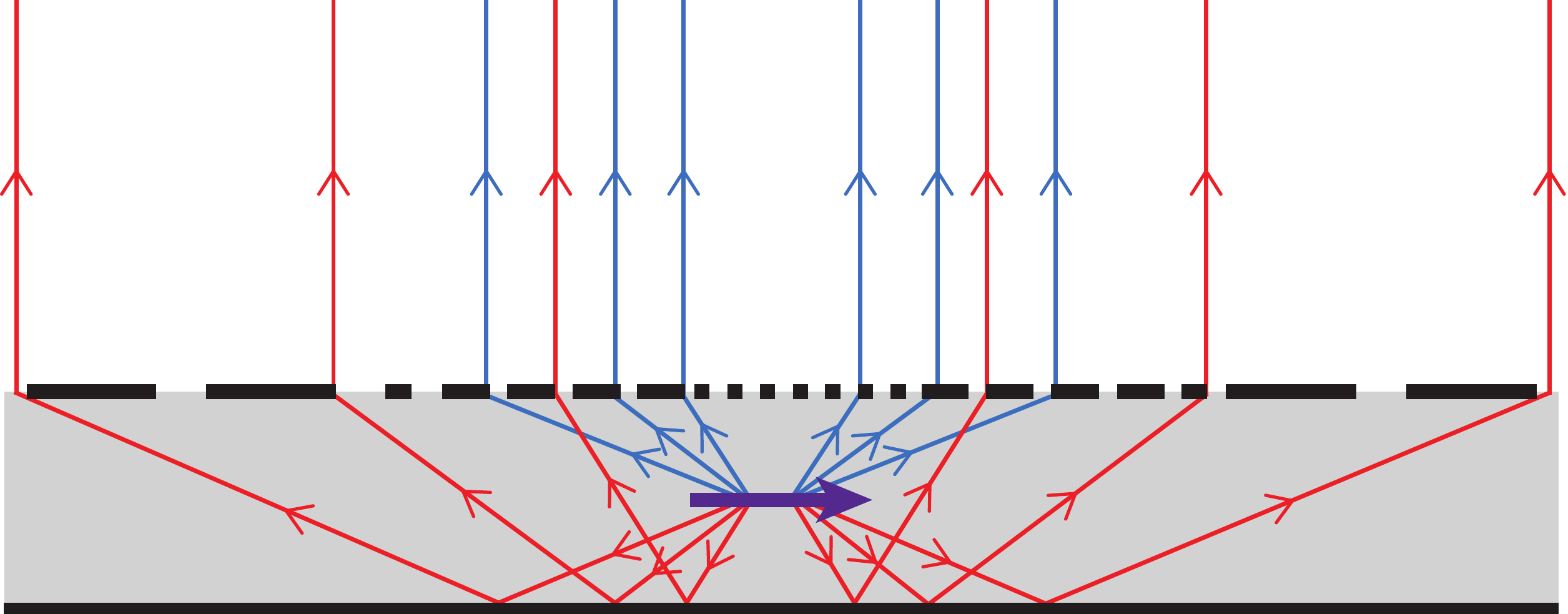}
\label{fig:led2}}\\
\subfloat[]{\includegraphics[width=0.5\linewidth]{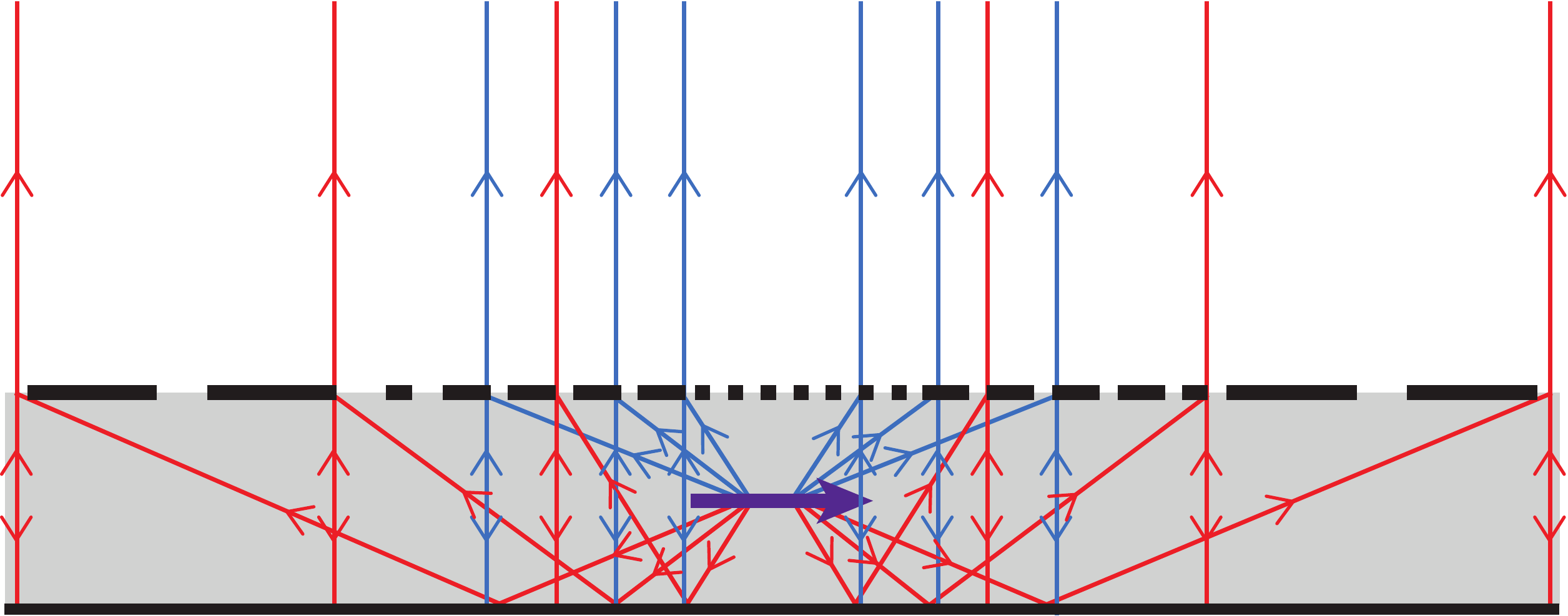}
\label{fig:led3}}
\subfloat[]{\includegraphics[width=0.5\linewidth]{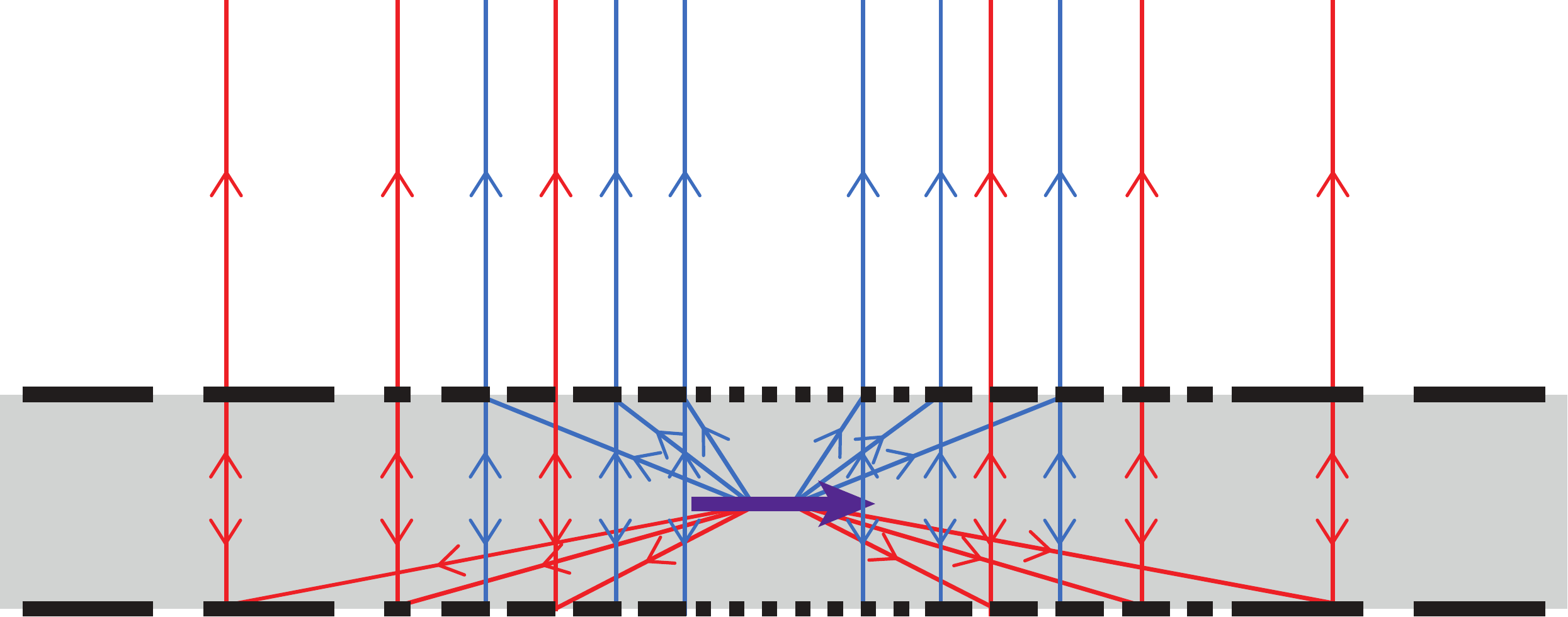}
\label{fig:led4}}
\caption{Radiation of a light source (quantum well) embedded in a semiconductor (e.g. GaN) substrate. (a)~Bare structure. (b)~Reflectionless metasurface, placed on top of the slab, that collimates the dipole fields. (c)~Introduction of perfectly refractive metasurface cavity (PRMC). (d)~Double metasurface cavity, with partially reflective top metasuface and fully reflective bottom metasuraface.}\label{fig:led}
\end{figure}

The LEE is enhanced by enforcing the emitted light to optimally refract/radiate perpendicularly to the device. Such refraction suppresses the wave trapping loss, represented in Fig.~\ref{fig:led1}. The requirement of total normal refraction, represented in Fig.~\ref{fig:led2}, is excessively stringent, leading to susceptibilities with prohibitive spatial variations, and is not required in this application. A better strategy consists, as illustrated in Fig.~\ref{fig:led3}, in allowing partial local reflection, and ultimately collecting the reflected part of the energy by Fabry-Perot resonance in the PRMC formed with a mirror plane at the bottom of the slab. The double-metasurface cavity, depicted in Fig.~\eqref{fig:led4} is an even more sophisticated design, leading to dramatic LEE enhancement.

The SER is enhanced by maximizing the confinement of coherent electromagnetic energy in the vicinity of the source and leveraging the Purcell effect, which is particularly well achieved in the double-metasurface PRMC (Fig.~\ref{fig:led4}). Finally, the far-field directivity is maximized as an optimization tradeoff for maximal overall power conversion ratio. 

\begin{figure}[htbp]
\begin{center}
\includegraphics[width=1\columnwidth]{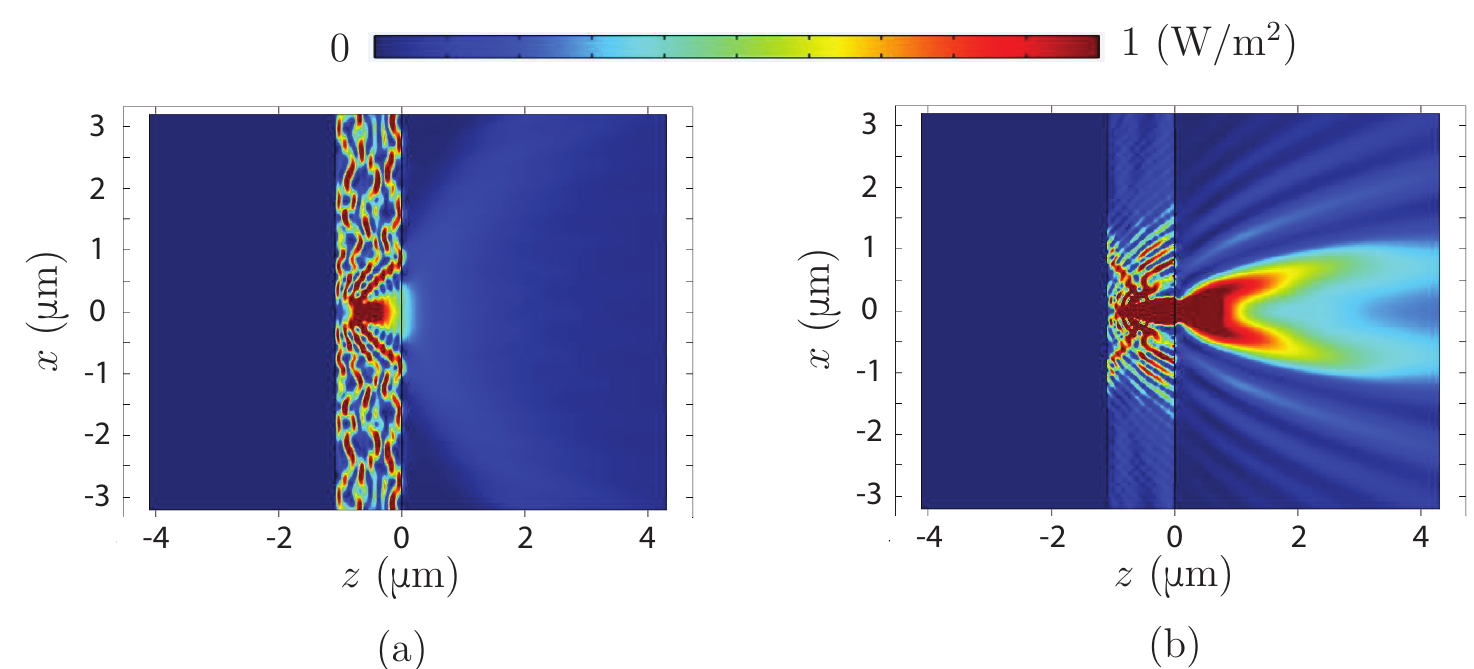}
\caption{Full-wave (COMSOL) simulated energy flux densities for a dipole emitter embedded in a GaN slab. (a)~Configuration of Fig.~\ref{fig:led1}. (b)~Configuration of Fig.~\ref{fig:led4}. Original images from~\cite{Chen2016}.}
\label{Fig:LEDsim}
\end{center}
\end{figure}

Figure~\ref{Fig:LEDsim} shows full-wave simulated flux densities for the designs of Figs.~\ref{fig:led1} and~\ref{fig:led4}, where the latter features LEE and SER enhancements by factors of 4.0 and 1.9, respectively, with half-power beam
width of 22.5$^\circ$. 
The case of a real LED is more complex due to the incoherence and distribution emission of the quantum well emitters. Different metasurface strategies are currently being investigated to maximize the power conversion efficiency of a complete LED.

\subsection{Second-Order Nonlinearity}

So far, we have only discussed linear metasurfaces, i.e. metasurfaces whose polarization densities are linear functions of the electric and magnetic fields. Given the wealth of potential applications of nonlinear metasurfaces, it is highly desirable to develop tools for the design of such metasurfaces. Therefore, we extended our susceptibility-GSTC technique to the case of a second-order nonlinear metasurface in~\cite{achouri2017mathematical}.

In this case, the polarization densities can be written as
\begin{subequations}\label{eq:NLPol}
\begin{align}
\ve{P} &= \epsilon_0\te{\chi}^{(1)}_\text{ee}\cdot\ve{E}_\text{av} + \epsilon_0\te{\chi}^{(2)}_\text{ee}:\ve{E}_\text{av}\ve{E}_\text{av},\\
\ve{M} &= \te{\chi}^{(1)}_\text{mm}\cdot\ve{H}_\text{av} + \te{\chi}^{(2)}_\text{mm}:\ve{H}_\text{av}\ve{H}_\text{av},
\end{align}
\end{subequations}
where $\te{\chi}^{(1)}$ and $\te{\chi}^{(2)}$ are to the linear and nonlinear (second-order) susceptibilities of the metasurface. For the sake of simplicity, we assume that these susceptibility tensors are scalar. Being nonlinear, the metasurface will generate harmonics of the excitation frequency $\omega_0$. Consequently, we have to express the GSTCs in~\eqref{eq:BC} in the time-domain to properly take into account the generation of these new frequencies. The relevant GSTCs are then, in the case of $x$-polarized waves, given by\footnote{In these expressions, the susceptibilities are dispersion-less. Meaning that $\chi(\omega_0)=\chi(2\omega_0)=\chi(3\omega_0)=...$, as discussed in~\cite{achouri2017mathematical}, which is essentially equivalent to the conventional condition of phase-matching in nonlinear optics.}
\begin{subequations}\label{eq:TDgstc}
\begin{align}
-\Delta H &= \epsilon_0 \chi^{(1)}_\text{ee} \frac{\partial}{\partial t} E_\text{av} + \epsilon_0 \chi^{(2)}_\text{ee} \frac{\partial}{\partial t} E^2_\text{av},\\
-\Delta E &= \mu_0 \chi^{(1)}_\text{mm} \frac{\partial}{\partial t} H_\text{av} + \mu_0 \chi^{(2)}_\text{mm} \frac{\partial}{\partial t} H^2_\text{av}\label{eq:TDgstc2},
\end{align}
\end{subequations}
where $E$ and $H$ are, respectively, the $x$-component of the electric field and the $y$-component of the magnetic field. From these relations, we can either perform a synthesis, i.e. expressing the susceptibilities as functions of the fields, or an analysis, i.e. computing the fields scattered from a metasurface with known susceptibilities. Here, for the sake of briefness, we will not elaborate on the synthesis and analysis operations but shall rather present one of the main results obtained in~\cite{achouri2017mathematical}, which are the reflectionless conditions for the metasurface. The metasurface with susceptibilities~\eqref{eq:TDgstc} exhibit different reflectionless conditions for the two propagation directions since, due to the presence of the square of both the electric and magnetic fields, the relations~\eqref{eq:TDgstc} are asymmetric with respect to the $z$-direction. It follows that the reflectionless conditions for waves propagating in the forward (+$z$) direction are
\begin{subequations}\label{eq:RLCond}
\begin{align}
\chi^{(1)}_\text{ee} = \chi^{(1)}_\text{mm},\\
\eta_0\chi^{(2)}_\text{ee} = \chi^{(2)}_\text{mm},
\end{align}
\end{subequations}
while for backward (-$z$) propagation they are
\begin{subequations}\label{eq:RLCond2}
\begin{align}
\chi^{(1)}_\text{ee} = \chi^{(1)}_\text{mm},\\
-\eta_0\chi^{(2)}_\text{ee} = \chi^{(2)}_\text{mm}.
\end{align}
\end{subequations}
An important consequences of the fact that the metasurface cannot be matched from both sides is that its second-harmonic generation (SHG) is inherently nonreciprocal.

\section{Conclusions}

We have presented an overview of electromagnetic metasurface designs, concepts and applications based on a bianisotropic surface susceptibility tensors model. This overview probably represents only a small fraction of this approach, which nevertheless already represents a solid foundation for future metasurface technology.

\section*{Acknowledgment}

This work was accomplished in the framework of the Collaborative Research and Development Project CRDPJ 478303-14 of the Natural Sciences and Engineering Research Council of Canada (NSERC) in partnership with the company Metamaterial Technology Inc.

\bibliographystyle{myIEEEtran}
\bibliography{NewLib}

\end{document}